\documentclass{ametsocV6}

\nolinenumbers

\title{How do cold pools influence the size of tropical cyclone embryos?}

\authors{Hao Fu\aff{a}\aff{b}\aff{c}\correspondingauthor{Hao Fu, haofu@nju.edu.cn. } 
}

\affiliation{
\aff{a}{School of the Atmospheric Sciences, Nanjing University, Jiangsu, China}\\
\aff{b}{Department of the Geophysical Sciences, University of Chicago, Illinois}\\
\aff{c}{Department of Earth System Science, Stanford University, California}\\
}

\abstract{The size of tropical cyclone (TC) embryos is an essential predictor of TC genesis. Recent studies have identified cold pools and planetary rotation as factors that increase and decrease TC embryo size. While the planetary rotation effect has been depicted using a quasi-geostrophic (QG) model, the cold pool effect still lacks a theoretical model. This paper presents a cloud chain model to derive the length scale regarding the influence of cold pools on the TC embryo vortex. Within the model, the amount of rain evaporation during a single convective event determines the wind speed and humidity at the cold pool edge, influencing the amount of sub-cloud moisture convergence for the next-generation convection and, therefore, the intensity of the next-generation cold pool. A perturbation analysis shows that cold pools exhibit a nonlocal dependence on air-column humidity, with the influence range determined by the cold pool size and a convective memory weight. The memory weight relies on the sum of the contributions of mechanical lifting and thermodynamic forcing to convective initiation. A crucial parameter is the ratio of rain evaporation to surface evaporation in a cold pool. By coupling the cloud chain model with the QG equation, an analytical expression for the TC embryo size is obtained. The theory captures the trend but overestimates the TC embryo size in cloud-permitting simulations. The deviation might be due to the oversimplification in estimating the fractional contribution of cold pools to convective initiation.}
\begin{document}

%% Necessary!
\maketitle

%%%%%%%%%%%%%%%%%%%%%%%%%%%%%%%%%%%%%%%%%%%%%%%%%%%%%%%%%%%%%%%%%%%%%
% SIGNIFICANCE STATEMENT/CAPSULE SUMMARY
%%%%%%%%%%%%%%%%%%%%%%%%%%%%%%%%%%%%%%%%%%%%%%%%%%%%%%%%%%%%%%%%%%%%%
%
% If you are including an optional significance statement for a journal article or a required capsule summary for BAMS 
% (see www.ametsoc.org/ams/index.cfm/publications/authors/journal-and-bams-authors/formatting-and-manuscript-components for details), 
% please apply the necessary command as shown below:
%
% Significance Statement (all journals except BAMS)
%
% \begin{sig}
% In the tropics, deep convective clouds interact with one another, spreading the convective activity to neighboring areas. Nevertheless, how to determine the diffusivity remains unclear. This paper derives eddy diffusivity and a cold pool transport length scale from fundamental cold pool dynamics. The new theory provides a reference for understanding what determines the size of a tropical convective system, such as a tropical cyclone precursor vortex. 
% \end{sig}
%
%% Capsule (BAMS only)
%%
%\begin{capsule}
%       Enter BAMS capsule here, no more than 30 words. See \url{www.ametsoc.org/index.cfm/ams/publications/author-information/formatting-and-manuscript-components/#capsule} for details.
%\end{capsule}
% 
%% \uparrow \uparrow If using twocol mode, you will need to use the commands "twocolsig" and "twocolcapsule" in place of "sig" and "capsule"
%%      to ensure that the text box correctly spans across both columns.

%%%%%%%%%%%%%%%%%%%%%%%%%%%%%%%%%%%%%%%%%%%%%%%%%%%%%%%%%%%%%%%%%%%%%
% MAIN BODY OF PAPER
%%%%%%%%%%%%%%%%%%%%%%%%%%%%%%%%%%%%%%%%%%%%%%%%%%%%%%%%%%%%%%%%%%%%%

\section{Introduction}

The mechanism of tropical cyclone (TC) genesis remains not fully understood \cite[e.g.,][]{montgomery2016recent,emanuel2018review,tang2020review}. A TC embryo refers to an early-stage vortex capable of developing into a TC but still lacks a prominent surface circulation. That is to say, the vortex is still before a tropical depression stage and cannot be explained by classic theories: the Ekman circulation is too weak to converge moisture effectively, and the surface wind is too weak to locally enhance air-sea heat exchange \citep{charney1964growth,emanuel1986air,Rotunno_and_Emanuel_1987}. Theoretical modeling of TC embryos is challenging, likely due to their asymmetric vortex structure and scattered convective activity \citep{ooyama1982conceptual,montgomery2006vortical}. Recently, a two-part paper series \citep{fu2023small,fu2024small_part2} developed a linear stability analysis framework to solve for the size and growth rate of TC embryos. Identifying the moisture-radiation feedback \cite[e.g.,][]{gray1977diurnal,wing2016role,ruppert2020critical} as the key mechanism of instability growth, they found the most unstable wavelength is clipped between the long- and short-wavelength cutoff. The long-wavelength cutoff is imposed by planetary rotation, which enforces the overturning circulation to be within a Rossby deformation radius \citep{fu2024small_part2}. A greater Coriolis parameter $f$ reduces the Rossby deformation radius and reduces the vortex size. The short-wavelength cutoff has been attributed to two effects \citep{fu2023small}. One is the cloud anvil's nonlocal radiative heating, which laterally extends beyond the cloud updraft \citep{adames2016mjo}. The other is the diffusion of convective activity by cold pools, which still lacks a theoretical model. The key challenge is that there is a scale separation between the size of a cold pool and the vortex overturning circulation. To depict the numerous clouds and cold pools, a statistical-mechanics perspective of the cloud population is necessary, which is the topic of this paper.

Deep convective towers interact via cold pools in the boundary layer \cite[e.g.,][]{tompkins2001organization,droegemeier1985three,feingold2013coupled,moseley2016intensification,grant2018role,torri2019cold}, as well as gravity waves \citep[e.g.,][]{mapes1993gregarious,liu2004vertical_mode,lane2011coupling,sakaeda2023observed} and stratified turbulence \cite[e.g.,][]{lilly1983stratified,fu2024stochastic} in the free troposphere. Cloud interaction has been shown to shape many statistical properties of quasi-equilibrium convection, such as the mean cold pool size and mean updraft size \citep{boing2012influence,schlemmer2014formation,nissen2021circling,torri2019cold,fu2025mean_part1,fu2025mean_part2}. 

\begin{figure}[h]
 \centerline{\includegraphics[width=0.7\linewidth]{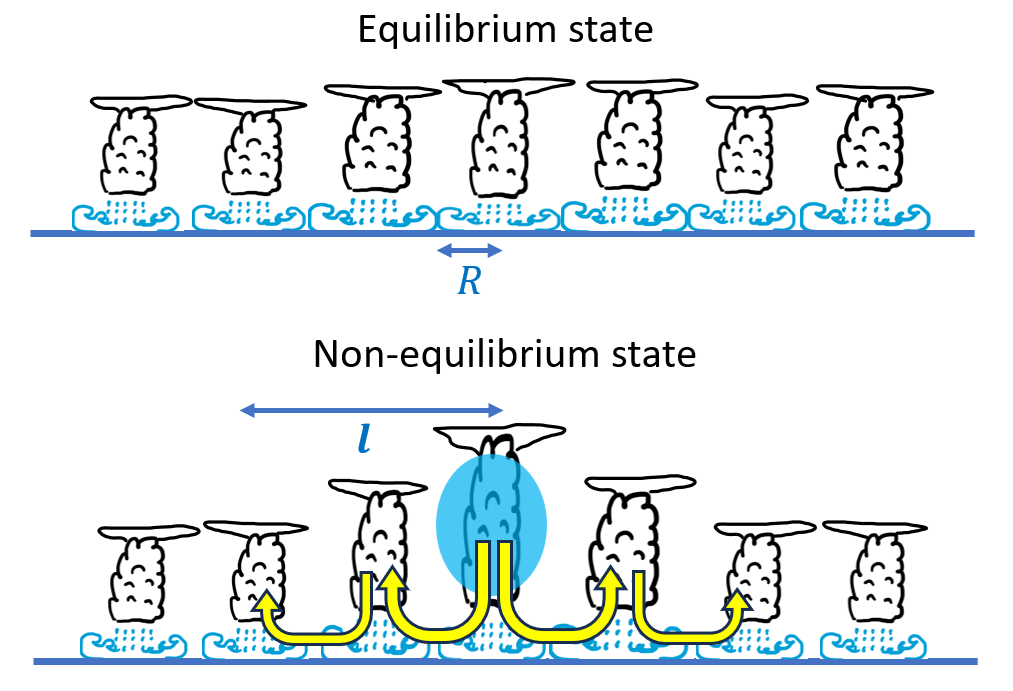}}
  \caption{A schematic diagram for the diffusion of convective activity induced by cold pools. Suppose there is an array of clouds. The mean radius of cold pools is ${R}$. The upper panel shows an equilibrium state where the PW and convective activity are horizontally homogeneous. The lower panel shows a non-equilibrium state in which the overturning circulation of a TC embryo produces a locally moist air column, denoted as the blue shading. Water vapor is transported laterally by cold pools, influenced by entrainment and surface evaporation, affecting neighboring convection. This influencing chain is marked with yellow arrows. This paper studies the influencing length scale of a humid air column on the distant convective/cold pool activity, ${l}$, which depends on $R$ and a ``convective memory weight" $\mathcal{W}$.} \label{fig:cartoon_spread}
\end{figure}

When there is background circulation or a sea surface temperature gradient, deviations from equilibrium emerge \citep{bretherton2005energy,sobel2006boundary,nolan2016itcz,skyllingstad2019modeling}. At different horizontal locations, the vigor of convective activity and the humidity of the air column differ. Thus, there could be transport processes that counteract the gradient \cite[e.g.,][]{craig2013coarsening,hottovy2015spatiotemporal,windmiller2019universality,ahmed2019stochastic,shi2021CSAdiffusion}. A classic example of the transport process is molecular diffusion, where gas molecules transport energy and momentum between each other via collision \cite[e.g.,][]{batchelor2000book}. Similarly, could cloud interaction be an \textit{effective} diffusive process that spontaneously reduces the gradient of air column humidity? This paper focuses on the cloud interaction via cold pools in the context of TC embryos, as illustrated in Fig. \ref{fig:cartoon_spread}. Previous studies provide two perspectives: the transport of water vapor by cold pools and the propagation of convective dynamical properties.

From the first perspective, cold pools spread convective activity by laterally transporting water vapor in the mixed layer. \citet{jeevanjee2013convective} found that cold pools suppress convective self-aggregation because they mix water vapor between the relatively moist and dry regions. \citet{jensen2022diurnal} used particle tracking to confirm that cold pools induce a dispersion effect on boundary layer tracers, and long-distance transport is inefficient due to the random occurrences of new cold pools. However, caution should be exercised when studying the ``lateral transport of water vapor", because rain evaporation, vertical advection by downdrafts, and wind-induced surface evaporation involve vertical transport and phase change that significantly modulate the transport of humidity in the mixed layer \citep{raymond1995BLQE,de2018variations}. 

The second perspective considers the dynamics of updrafts, downdrafts, and cold pools as an entity: the convective lifecycle \cite[e.g.,][]{byers1949thunderstorm,ferrier1989one,feng2015cold_pool}, and studies how these properties propagate. The convective lifecycle has a significant influence on the ``convective memory", a terminology devised to conceptualize the dependence of convective activity on its history, irrespective of the instantaneous environmental state \citep{cohen2004response,davies2009memory,davies2013departures,yano2012ODEs,colin2019identifying,colin2021atmospheric}. Cold pools produced by previous convection collide with each other, initiating new convection in the neighborhood of the previous convection \cite[e.g.,][]{tompkins2001organization,haerter2019circling,torri2019cold,haerter2020diurnal,yang2021parcel}. Convective initiation by cold pools is mainly attributed to mechanical lifting and thermodynamic forcing \citep{falk2022environmental}. 
\begin{itemize}
    \item Mechanical lifting process denotes the lifting of air parcels to the level of free convection against convective inhibition energy \citep{droegemeier1985three,grandpeix2010density,meyer2020mechanical}. 
    \item Thermodynamic forcing process denotes the aggregation of water vapor at the cold pool's edge (gust front), which humidifies the mixed layer top upon collision and provides fuel for convective initiation \citep{tompkins2001organization,langhans2015ring,fuglestvedt2020cold}. The moisture source includes rain evaporation during the previous convection and the wind-induced surface evaporation. 
\end{itemize}
Cold pools have been shown to contribute significantly to convective memory \citep{colin2019identifying}, and how to model this process quantitatively is the next challenge. The phenomenological lattice model of \citet{haerter2019convective} depicted the spread of convective activity as a lateral diffusion process. A more detailed investigation was performed by \citet{yang2021parcel}, who coupled an array of Lagrangian air parcel cloud models with simple cold pool dynamics. The cold pool speed is set to a constant, and the mechanical lifting effect of a cold pool is quantified as a constant velocity scale. Cold pools are observed to spread convective activity in their model. This is an enlightening model. Nevertheless, we argue that the constant assumption neglects some key dynamics of cold pools and could not answer the relative role of mechanical lifting and thermodynamic forcing in the convective spreading behavior.

This paper proposes a quantitative model to reconcile the two perspectives: the convective triggering chain. It considers both convective and cold pool processes in cloud interaction, and incorporates the transport of water vapor and the propagation of convective dynamical properties. We split the convective lifecycle into two phases: the convective phase and the cold pool phase. Different from \citet{yang2021parcel}, who carefully modeled the convective phase and treated cold pools relatively crudely, we focus on cold pools and treat the convective phase crudely, an approach possibly more suitable for this specific scientific question. For the convective phase, we assume the sub-cloud rain evaporation amount is proportional to the sub-cloud moisture convergence. The proportional factor is related to the air-column humidity. For the cold pool phase, we quantify the sub-cloud moisture convergence rate as the product of a sub-cloud convergence wind speed and the humidity at the gust front (cold pool edge). Sub-cloud convergence is partitioned between the cold pool and the mixed-layer convective cells. Because the gust front speed is directly linked to its potential temperature anomaly, the problem reduces to modeling the gust front's potential temperature and moisture evolution. These two thermodynamic quantities depend on the initial rain evaporation amount at the cold pool formation stage and the surface flux at the cold pool propagation stage. The rain evaporation during the cold pool formation stage carries convective memory from neighboring clouds. This enables us to couple the clouds into a chain model that permits an analytical solution in the weakly non-equilibrium case. We aim to address the nonlocal influence length scale due to cold pools, $l$.

As for the organization of this paper, section \ref{sec:CM1} uses cloud-permitting simulations with varying sub-cloud evaporation rates to demonstrate how cold pools influence the size of simulated TC embryos. Section \ref{sec:theory} presents the cloud chain model and analytically solves the nonlocal influence length scale $l$. Section \ref{sec:validation} couples the cloud chain model with the QG model to solve for the TC embryo size and benchmarks it with cloud-permitting simulations. Section \ref{sec:discussion} discusses the limitations of the theory, and section \ref{sec:conclusion_wavepacket} concludes the paper.

\section{Cloud-permitting simulations} \label{sec:CM1}

\subsection{The spontaneous TC genesis problem}

This section uses cloud-permitting simulations to diagnose how cold pools influence the size of TC embryos. An idealized experimental setup that permits inhomogeneity in the convective activity is needed. The inhomogeneity can be implemented in the boundary condition, such as a nonuniform sea surface temperature \cite[SST, e.g.,][]{skyllingstad2019modeling}, or can be produced spontaneously by internal feedbacks of the atmosphere, such as the moisture-radiation feedback \cite[e.g.,][]{ruppert2020critical}. We choose the latter and study the problem in the context of spontaneous TC genesis. 

Spontaneous TC genesis denotes the emergence of TC-like vortices in a numerical setup with a constant Coriolis parameter and a uniform SST \cite[e.g.,][]{bretherton2005energy,khairoutdinov2013rotating,davis2015formation,wing2016role,carstens2022f}. The early stage of spontaneous TC genesis features a group of 100 km scale mesoscale vortices, which are TC embryos. Each vortex hosts several deep convective towers at a time. Later on, some embryo vortices merge and some dissipate, until a few mature TCs form in the domain. \citet{fu2023small} introduced the technique of amplifying the longwave radiative feedback to enhance the signal of these TC embryos, which would otherwise be hard to distinguish from noise.

\subsection{Experimental setup}

The simulations are performed with Bryan Cloud Model 1 \cite[CM1 v19.10,][]{bryan2002benchmark}, using a doubly periodic domain with a width of 1080 km by 1080 km and a horizontally uniform grid spacing of 2 km. The domain top is at $z=27$ km, with a sponge layer above $z=20$ km to absorb gravity waves. Among the 65 vertical mesh layers, eight layers are deployed below 1 km height to resolve cold pools better. The initial background wind speed is prescribed to be zero. The SST is set to a fixed value of 27 $^\circ$C. The model uses the \citet{morrison2005new} double moment microphysical scheme, the RRTMG radiative transfer scheme without considering the diurnal cycle \citep{clough2005RRTM}, the boundary layer scheme developed by \citet{bryan2009maximum}, and the surface flux scheme developed by \citet{jimenez2012revised}. The simulations are initialized with a radiative-convective equilibrium background vertical profile generated by a small-domain simulation (with a width of 120 km, and no Coriolis force, all other parameters remaining the same) plus a random potential temperature perturbation with a magnitude of 0.1 $^\circ$C in the lowest five levels of the model. 

%The Coriolis parameter uses $f=10^{-4}$ s$^{-1}$, corresponding to an effective latitude of $43.3^\circ$N. 

Two groups of experiments with a total of 19 simulations are conducted, as listed in Table \ref{Table_experiments}. One group tests twelve values of the sub-cloud rain evaporation rate. The other group tests eight values of the Coriolis parameters from $0.25\times10^{-4}$ s$^{-1}$ to $2\times10^{-4}$ s$^{-1}$. They share one reference experiment (normal rain evaporation rate and $f=10^{-4}$ s$^{-1}$). The two groups will reveal the short- and long-wavelength cutoff factors of TC embryo size. Most calculations use data at $t=2.5$ days, by which time the TC embryos are weak but identifiable. 

\begin{table}[h]
\caption{The parameters of the nineteen CM1 simulations that change the sub-cloud rain evaporation rates (by multiplying the $\mathrm{E_v}$ factor) and change the Coriolis parameter $f$. }\label{Table_experiments}
\begin{center}
\begin{tabular}{ccccrrcrc}
\hline
\hline
Name  &  $\mathrm{E_v}$ &  $f$ (s$^{-1}$) \\
\hline
Ev-0.2-f-1 & 0.2 &  $1\times10^{-4}$  \\
Ev-0.5-f-1 & 0.5 &  $1\times10^{-4}$  \\
Ev-0.7-f-1 & 0.7 &  $1\times10^{-4}$  \\
Ev-1.0-f-1 (reference) & 1.0 &  $1\times10^{-4}$  \\
Ev-1.2-f-1 & 1.2 &  $1\times10^{-4}$  \\
Ev-1.5-f-1 & 1.5 &  $1\times10^{-4}$  \\
Ev-1.7-f-1 & 1.7 &  $1\times10^{-4}$  \\
Ev-2.0-f-1 & 2.0 &  $1\times10^{-4}$  \\
Ev-2.2-f-1 & 2.2 &  $1\times10^{-4}$  \\
Ev-2.5-f-1 & 2.5 &  $1\times10^{-4}$  \\
Ev-2.7-f-1 & 2.7 &  $1\times10^{-4}$  \\
Ev-3.0-f-1 & 3.0 &  $1\times10^{-4}$  \\
\hline
Ev-1-f-0.25 & 1.0 &  $0.25\times10^{-4}$  \\
Ev-1-f-0.5 & 1.0 &  $0.5\times10^{-4}$  \\
Ev-1-f-0.75 & 1.0 &  $0.75\times10^{-4}$  \\
Ev-1-f-1.0 (reference) & 1.0 &  $1.0\times10^{-4}$  \\
Ev-1-f-1.25 & 1.0 &  $1.25\times10^{-4}$  \\
Ev-1-f-1.5 & 1.0 &  $1.5\times10^{-4}$  \\
Ev-1-f-1.75 & 1.0 &  $1.75\times10^{-4}$  \\
Ev-1-f-2.0 & 1.0 &  $2.0\times10^{-4}$  \\
\hline
\hline
\end{tabular}
\end{center}
\end{table}

\subsection{More details on the experimental design}

First, we acknowledge that modifying rain evaporation rate is a widely used method for changing the intensity of cold pools or downdrafts in idealized simulations \cite[e.g.,][]{jeevanjee2013convective,grant2018role,yang2018length_agg,wang2019dynamics,rios2020impacts,nissen2021circling}. In our simulations, the rain evaporation rate variable (``EPSR" in ``morrison.F" script) is multiplied by a parameter $\mathrm{E_v}$ in the microphysics scheme within the lowest 500 m of the domain, confining the influence to the mixed layer.\footnote{We comment that this method changes the horizontally averaged rain evaporation rate as well, which modulates the domain climate. In section \ref{sec:validation}, we will demonstrate that a higher $\mathrm{E_v}$ increases the Bowen ratio by moistening the mixed layer. It also influences the buoyancy stratification across the domain by cooling the mixed layer and could affect the convective initiation process. } Twelve experiments from $\mathrm{E_v}=0.2$ to 3.0 are performed. A similar yet smaller set of experiments has been reported by \citet{fu2023small}, who found that stronger sub-cloud evaporation widens TC embryos. This paper digs into the physical mechanism.

Second, there are two reasons for choosing a relatively high Coriolis parameter of $f=10^{-4}$ s$^{-1}$ when studying the cold pool effects:
\begin{enumerate}
    \item A sufficiently large $f$ is needed to suppress inertial gravity waves in the domain \citep{fu2023small,fu2024small_part2}. The waves couple with convection through the vertical motion component, complicating the analysis of cold pools. Increasing $f$ also suppresses the growth of TC embryos, but less significantly than on waves \citep{fu2024small_part2}. 
    \item A higher $f$ reduces the size of simulated TC embryos \citep{fu2024small_part2} to only several times the cold pool size. This makes the spread of convective activity induced by cold pools more prominent and, therefore, easier to study. In addition, smaller vortices correspond to higher populations, improving the accuracy of vortex size diagnosis. 
\end{enumerate}
In the next subsection, experiments with varying $f$ will confirm that $f$ does not significantly influence the 10 km scale cold pool dynamics unless $f$ approaches $2\times10^{-4}$ s$^{-1}$.

Third, an essential modification to all experiments is that the horizontal anomaly of longwave radiative heating tendency has been doubled to make the mesoscale vortices grow faster and, therefore, easier to diagnose at the early stage. The stronger horizontal anomaly of longwave heating enhances the local greenhouse effect imposed by upper clouds, amplifying the circulation in the vertical cross-section and fostering moisture convergence. More details of this technique's implementation procedures and effects are documented in \citet{fu2023small}.\footnote{We have checked the $\mathrm{E_v}=1$ experiment and found that the shortwave plus longwave radiative heating rate in cold pool regions has a much smaller magnitude than the sensible heating rate (shown in Fig. S1 in the Zenodo dataset). The longwave radiative emission of cold pools seems to be largely canceled by the absorption of upward longwave radiance from the sea surface. Future work is needed on studying the role of radiative heating in cold pool dynamics, which has received little attention.} If the longwave radiative feedback is not amplified, vortices eventually form, as shown in many previous simulations \cite[e.g.,][]{davis2015formation,wing2016role,fu2023small}. However, once identifiable, there is usually only one or a few rapidly growing vortices, making comparison between experiments difficult. Thus, it is more straightforward to investigate the cloud interaction problem at the early stage, where multiple vortices emerge in an orderly pattern. Cold pools also play an important role at the late stage of TC genesis, such as through enhancing convective organization by the cold pool-shear interaction mechanism \citep{rotunno1988theory,davis2015formation}, as well as generating low-level vorticity perturbations \citep{smith2019creation}. This is a different topic left for future work.

\begin{figure}[h]
\centerline{\includegraphics[width=1.6\linewidth]{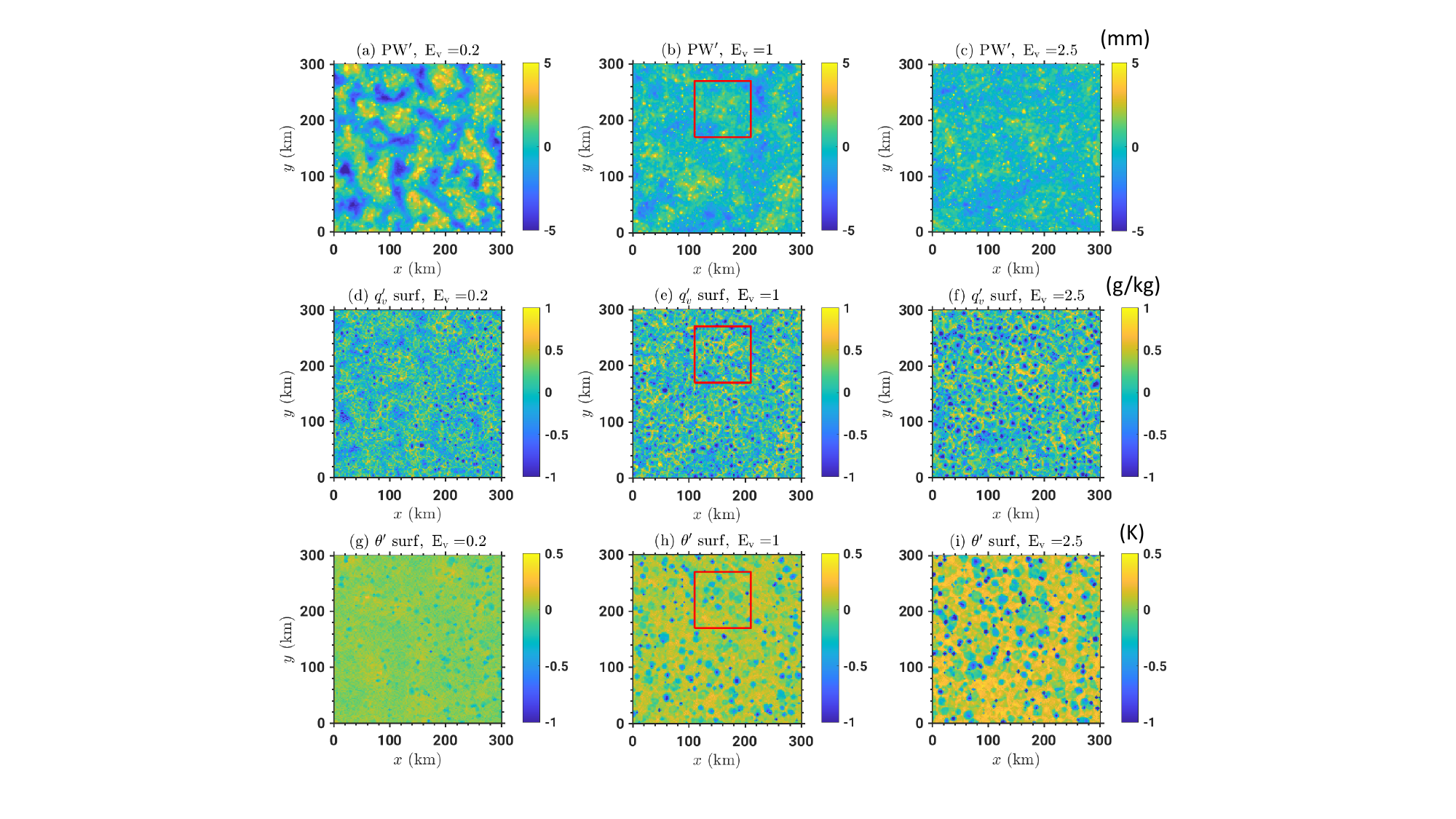}}
  \caption{Zoom-in plots of the 1080 km $\times$ 1080 km simulations to the 300 km $\times$ 300 km southwest corner region. The three columns show the $\mathrm{E_v}=0.2$, 1, and 2.5 experiments, respectively. All use $f=10^{-4}$ s$^{-1}$. The first row shows the horizontal anomaly of column precipitable water $\mathrm{PW}'$. The second row shows the horizontal anomaly of near-surface ($z=25$ m) water vapor mixing ratio $q'_s$. The third row shows the horizontal anomaly of the near-surface ($z=25$ m) potential temperature anomaly. The 100 km $\times$ 100 km red boxes highlight a moist patch in the $\mathrm{E_v}=1$ simulation. All plots use the data at the time $t=2.5$ days. A movie can be downloaded from: https://doi.org/10.5281/zenodo.17668821.}\label{fig:pcolor}
\end{figure}

\begin{figure}[h]
\centerline{\includegraphics[width=1.0\linewidth]{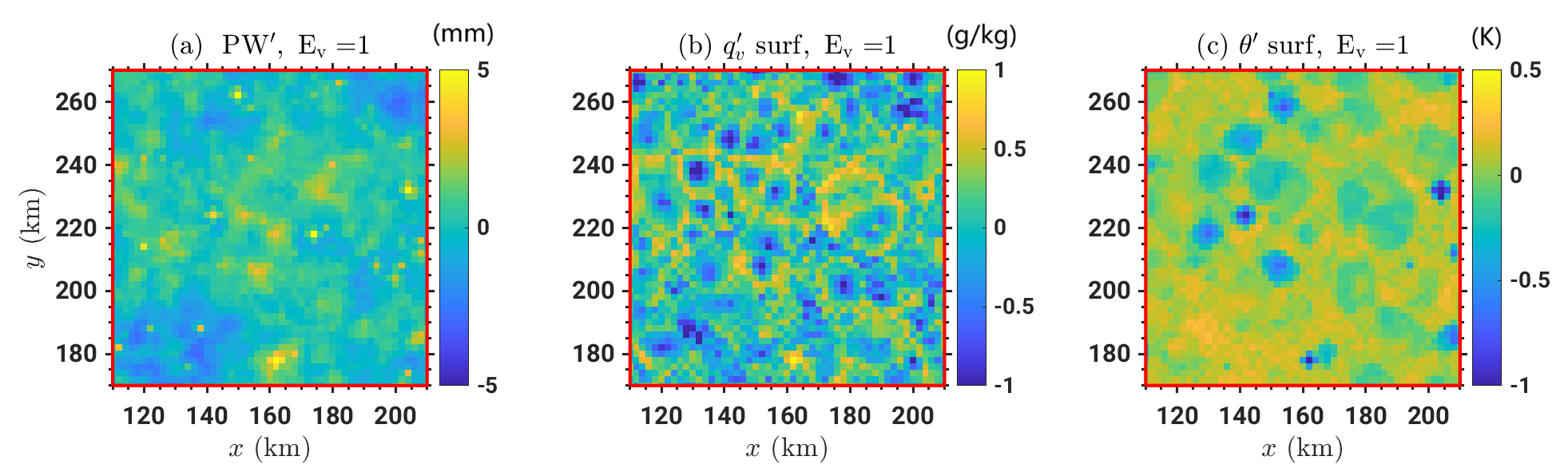}}
  \caption{A further zoom-in plot of the 100 km $\times$ 100 km red box region in Fig. \ref{fig:pcolor} for the $\mathrm{E_v}=1$ reference simulation. (a)-(c) show the horizontal anomaly of column precipitable water, near-surface water vapor mixing ratio, and near-surface potential temperature.  }\label{fig:pcolor_zoom}
\end{figure}

\subsection{Simulation results}

Figure \ref{fig:pcolor} shows that mesoscale vortices have spontaneously formed by $t=2.5$ days. The vortices correspond to regions with greater column precipitable water ($\mathrm{PW}$, unit: m). The PW denotes the thickness of a liquid water column if all the water vapor in the air column condenses to the surface, and is an integrated measure of humidity in the mixed layer and the free troposphere. This corresponding relation is consistent with the argument that a more humid region protects convection from entraining dry air and, therefore, fosters stronger convection and spinning up more substantial vortical circulations \cite[e.g.,][]{randall1980stochastic,derbyshire2004RH,raymond2007theory}. The 10 km-scale cellular structures reflected in the near-surface water vapor mixing ratio ($q_s$) are cold pools, which have a dry core produced by downdraft flow and a humid edge (gust front) produced by rain evaporation and surface evaporation \citep{tompkins2001organization,langhans2015ring,fu2025mean_part1,fu2025mean_part2}. See Fig. \ref{fig:pcolor_zoom} for a zoom-in plot. The cold pools are organized into 100 km-scale mesoscale structures, which correspond to TC embryo vortices. A significant trend is that higher $\mathrm{E_v}$ experiments show wider, weaker TC embryo vortices (Fig. \ref{fig:pcolor}). By weaker, we mean that the contrast of PW decreases. Next, we use a Fourier spectral analysis to quantify the multiscale structure. 

\begin{figure}[h] \centerline{\includegraphics[width=0.95\linewidth]{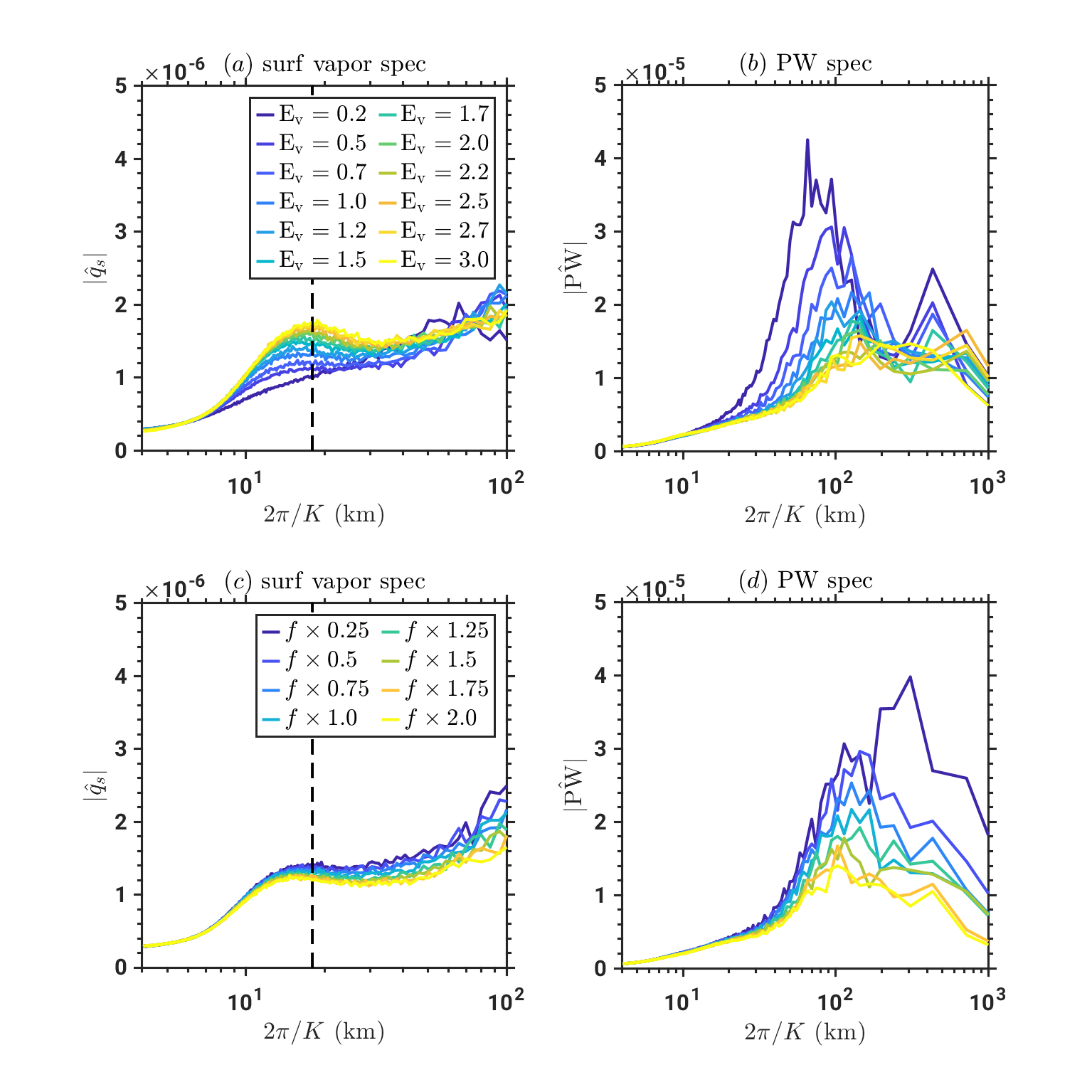}}
\caption{The modulus of the Fourier spectrum of (a) near-surface ($z=25$ m) water vapor mixing ratio $q_s$ and (b) column-precipitable water PW, averaged with 13 time snapshots between $t=2.5$ days$\pm$0.25 days. The first row shows experiments with varying $\mathrm{E_v}$, with the line color progressing from dark to bright indicating increasing $\mathrm{E_v}$. The second row shows experiments with varying $f$, with the line color progressing from dark to bright indicating increasing $f$. The dashed black lines in (a) and (c) show the 18 km wavelength as a reference. }   \label{fig:spectrum}
\end{figure}

The cold pool structure is analyzed with the Fourier transform of the near-surface water vapor mixing ratio, $\hat{q}_{s}$, which is expressed as: 
\begin{equation}   \label{eq_fourlayer:Fourier}
\hat{q}_{s} 
\equiv \frac{1}{L^2} \int_{0}^{L}\int_{0}^{L} q_s \exp\left[-i(k_x x+k_y y) \right] dxdy, 
\end{equation}
where $k_x$ and $k_y$ are horizontal wavenumbers, and $L=1080$ km is the width of the simulation domain. All spectra are plotted with $K \equiv \sqrt{k_x^2+k_y^2}$ as the abscissa, referred to as the horizontal total wavenumber. The spectrum of $|\hat{q}_{s}|$ is quite disordered above the 100 km scale, so we mainly use it to analyze the cold pool structure. Figure \ref{fig:spectrum}a shows that most simulations (except the weakest rain evaporation case $\mathrm{E_v}=0.2$) exhibit a spectral peak near 18 km wavelength, which informs the mean cold pool diameter. The $\mathrm{E_v}=0.2$ case does not have a characteristic cold pool diameter because the TC embryos develop sufficiently fast, making cold pool structures vanish in the relatively dry region (Fig. \ref{fig:pcolor}). The cold pool diameter in other experiments slightly increases with $\mathrm{E_v}$. In Appendix A, an alternative method is introduced to diagnose the mean cold pool radius from the decaying length scale of the spatial autocorrelation of $q_s$. This alternative method circumvents the influence from mesoscale structures like TC embryos, yielding a monotonic increase in cold pool radius with $\mathrm{E_v}$.

The mesoscale structure is analyzed with PW, a combined representation of mixed-layer and free-tropospheric humidity. For all simulations, the Fourier spectrum of PW, $\hat{\mathrm{PW}}$, exhibits a primary peak near 100 km scale, which informs the mean TC embryo size. In the first experimental group, as $\mathrm{E_v}$ increases, the peak shifts to longer wavelengths, indicating that more vigorous cold-pool activity widens TC embryo vortices. When changing $\mathrm{E_v}$, the shape of the spectrum mainly differs in the short-wavelength range, indicating that cold pools primarily influence the relatively small-scale dynamics. When $\mathrm{E_v}\lesssim 1$, a greater $\mathrm{E_v}$ induces a secondary peak at the long-wavelength range, which has been attributed to convectively coupled gravity waves by \citet{fu2023small}. The influence of sub-cloud rain evaporation on waves is left for future work. For the second experimental group, as $f$ increases, the peak shifts to shorter wavelengths, indicating that stronger planetary rotation narrows TC embryo vortices. When changing $f$, the spectrum mainly differs in the long-wavelength range, indicating that planetary rotation primarily influences the large-scale dynamics. As the short- and long-wavelength cutoffs approach, either due to stronger cold pools or stronger planetary rotation, the peak magnitude of $|\hat{\mathrm{PW}}|$ generally decreases, indicating weaker TC embryos. An exception is the $\mathrm{E_v}\gtrsim 2$ regime, where the TC embryo intensity shows a stagnant relationship with the cold pool intensity. This phenomenon is yet to be explained.

The long-wavelength cutoff has been explained as the effect of planetary rotation using the QG model of \citet{fu2024small_part2}. However, the short-wavelength cutoff has been explained only phenomenologically as a nonlocal dependence of the diabatic heating rate on column precipitable water. In the next section, we build a cloud-chain model to determine $l$ and then couple it with the QG model to obtain a theory of TC embryo size.

\section{The theoretical model}\label{sec:theory}

This section builds a 1-D cloud chain model to depict cloud interaction via cold pools. We introduce the model framework, then derive the governing equation, perform a perturbation analysis, and finally identify the nonlocal influence length scale. 

\subsection{Model framework} \label{subsec:framework}

\begin{figure}[h] \centerline{\includegraphics[width=0.85\linewidth]{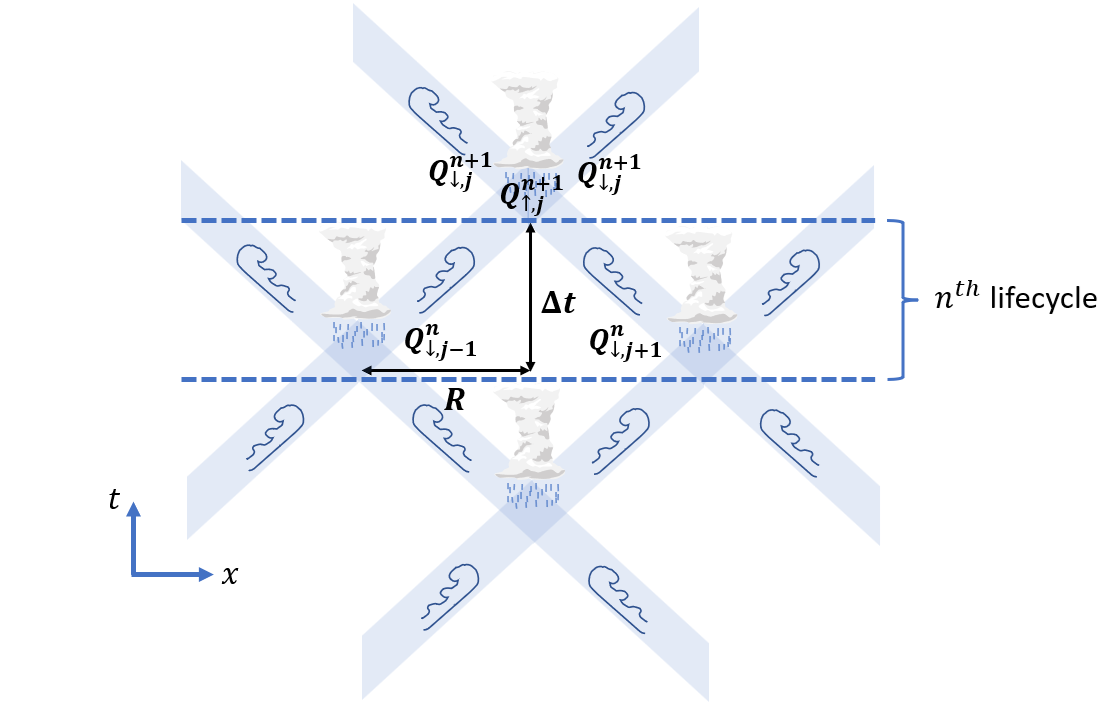}}
\caption{A schematic diagram of the 1-D cloud chain model, with the abscissa denoting the spatial dimension $x$ and the ordinate denoting time $t$. The blue shading shows the gust fronts' trajectories. The ${R}$ is the mean cold pool radius, which corresponds to the cold pool's propagation distance. The $\Delta t$ is the cold pool's propagation time. The slot between the two dashed lines denotes the $n^{th}$ lifecycle. The $Q^{n+1}_{\uparrow,j}$ denotes the latent heat transported to the updraft by sub-cloud moisture convergence at $x_j = j {R}$, and $Q^{n+1}_{\downarrow,j}$ denotes the sub-cloud latent heat absorption in the follow-up downdraft.  }   \label{fig:cartoon_hovmoller_wavepacket}
\end{figure}

The geometric assumptions are introduced first. The convective duration time \cite[less than an hour,][]{emanuel1994book} is much shorter than the cold pool propagation time \cite[around 2.5 hours,][]{tompkins2001organization}, and the updraft radius \cite[around 1 km,][]{fu2025mean_part2} is much smaller than the cloud spacing \cite[around 20 km,][]{fu2025mean_part2}. Assuming new convection is initiated by the collision of two cold pools, the cloud spacing is roughly twice the terminal radius of the cold pools, $R$. Cold pools in the simulations are at different lifecycle stages and therefore have a wide size distribution. For simplicity, the cloud spacing is taken as a constant. Thus, we introduce a 1-D array of clouds with a position index $j$ and a lifecycle number $n$, as illustrated in Fig. \ref{fig:cartoon_hovmoller_wavepacket}.\footnote{Using a cloud array to study cloud interaction was first proposed by \citet{bretherton1988field}. He studied the mutual suppression between nonprecipitating shallow convection via the drying effect of their descending motion. The evolution of convective strength and the movement of clouds are considered. \citet{feingold2013coupled} built a cloud array with fixed cloud spacing to study the phase synchronization of precipitating shallow convection. The influence of one cloud's height on the other via cold pools is considered phenomenologically using a coupled predator-prey model. 
Here, we study deep convection and explicitly consider the influence of free-tropospheric humidity, surface evaporation, and surface sensible heating. The focus is on the diffusion of convective activity rather than the phase synchronization.} The radius of a convective cloud is assumed to be infinitely small, and the duration of the cloud is assumed to be infinitely short. The location of each cloud is $x_j = j {R}$, and the time coordinate is $t = n \Delta t$. Here, $\Delta t$ is the propagation time of a cold pool before initiating the next-round convection. Appendix B extends the model to the more realistic 2-D geometry and shows that the main conclusion remains unchanged.

The convective lifecycle is assumed to be characterized by two quantities: the latent heat (water vapor) transported to the mixed layer top per gust front collision event, $Q^{n}_{\uparrow,j}$ (unit: J), and the latent heat absorbed per sub-cloud rain evaporation event, $Q^{n}_{\downarrow,j}$ (unit: J). The superscript $n$ denotes the $n^{th}$ lifecycle, with each cycle beginning at the moment of convective initiation. The subscript $j$ denotes the convective event at $x_j = j {R}$. Symbols with ``$\uparrow$" and ``$\downarrow$" denote quantities associated with convective initiation and cold pool formation, respectively. To build a model of cloud interaction, we need to understand the two phases of a convective lifecycle:
\begin{itemize}
    \item The cold pool phase - how does $Q^{n}_{\downarrow,j\pm1}$ determine $Q^{n+1}_{\uparrow,j}$? The key question is how the present cycle's sub-cloud evaporation amount determines the moisture that converges into the next cycle's updraft.  
    \item The convective phase - how does $Q^{n}_{\uparrow,j}$ determine $Q^{n}_{\downarrow,j}$? The key question is how moisture convergence into the updraft determines the amount of sub-cloud evaporation in the following downdraft. 
    \end{itemize}
Two steps are involved in the convective phase. First, the lifted water vapor condenses into rainwater, with a portion lost to the environment through dry-air entrainment. This conversion fraction increases with midlevel environmental humidity \citep{langhans2015lagrangian,lutsko2018increase,fu2019kinematic}. Second, some rainwater re-evaporates in the mixed layer. The re-evaporation rate increases with sub-cloud humidity \citep{torri2016lagrangian}. They indicate that competing factors control the dependence of cold pool initial intensity on PW. \citet{james2010numerical} showed that for a low-CAPE regime like the tropics, the first effect dominates. Thus, $Q^n_{\downarrow,j}$ not only represents the cold pool initial intensity, but also the rain formation amount in a convective event and will be linked to convective diabatic heating in section \ref{sec:validation}. The ratio of sub-cloud rain-evaporation amount to sub-cloud moisture convergence is denoted as $\eta$, which is set as a monotonically increasing function of the PW anomaly at the position of the $j^{th}$ cloud ($\mathrm{PW}'_j$):
    \begin{equation}   \label{eq:eta}
    Q^{n}_{\downarrow,j} 
    = \underbrace{ \eta_* \left( 1 + \frac{\mathrm{PW'}_j}{\mathrm{PW}_*} \right) }_{\eta}
    Q^{n}_{\uparrow,j}.    
    \end{equation}
The constant quantity $\eta_*$ is the $\eta$ when $\mathrm{PW'}_j=0$, and $\mathrm{PW}_*$ is a characteristic PW representing the sensitivity of $\eta$ to $\mathrm{PW'}_j$. When the air column is close to saturation, i.e., at the late stage of TC genesis, a moister air column may inhibit downdrafts \citep{emanuel1989finite}. This regime is not considered here. 

The duration of each cold pool phase is $\Delta t$, and that of each convective phase is neglected. Next, we focus on the cold pool phase.

\subsection{The cold pool dynamics}
\subsubsection{Convective initiation}

To study how $Q^{n+1}_{\uparrow,j}$ depends on $Q^{n}_{\downarrow,j \pm 1}$, we first investigate how $Q^{n+1}_{\uparrow,j}$ depends on the cold pool quantities. We let $Q^{n+1}_{\uparrow,j}$ be the latent heat amount transported to the top of the mixed layer in a cold pool collision event. As shown by \citet{tompkins2001organization}, for tropical convection, the air at the cold pool edge can directly become part of the updraft, so we express $Q^{n+1}_{\uparrow,j}$ as the outcome of moisture convergence. 

\begin{figure}[h] \centerline{\includegraphics[width=0.5\linewidth]{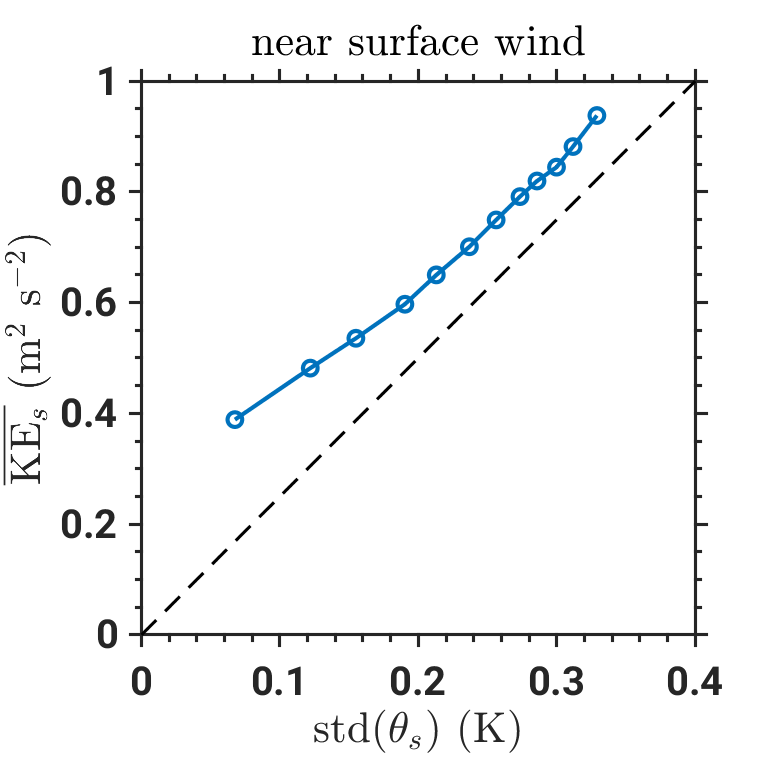}}
\caption{The circle line shows the standard deviation of near-surface potential temperature ($\theta_s$) versus the horizontally averaged near-surface kinetic energy of the irrotational wind ($\overline{\mathrm{KE}_s}$), using the twelve experiments with varying $\mathrm{E_v}$. The dashed line shows a reference: $\overline{\mathrm{KE}_s} = 2.5\, \mathrm{std}(\theta_s)$. The irrotational wind is diagnosed with the Helmholtz velocity decomposition. A Poisson equation with the horizontal divergence as the source term is inverted. All quantities use the cross-sectional data at $z=25$ m. }   \label{fig:theta_KE}
\end{figure}

How much of the convergent flow is due to cold pools, and how much is due to convective cells in the mixed layer? If the convergent flow were due to cold pools, one would expect that the magnitude of the near-surface wind speed, particularly its irrotational component, is proportional to the square root of the near-surface potential temperature anomaly \cite[e.g.,][]{torri2019cold}. Figure \ref{fig:theta_KE} diagnoses their relation at $t=2.5$ days for experiments with varying rain evaporation rates. The near-surface mean kinetic energy of the irrotational wind ($\overline{\mathrm{KE}_s}$) generally increases with the standard deviation of near-surface potential temperature [$\mathrm{std}(\theta_s)$]. When $\mathrm{E_v} \gtrsim 1$, they are proportional, indicating that the near-surface irrotational wind is mostly associated with cold pools. When $\mathrm{E_v} \ll 1$, $\overline{\mathrm{KE}_s}$ gradually gets insensitive to $\mathrm{std}(\theta_s)$, indicating that the near-surface irrotational wind is mostly contributed by convective cells. We let $u^n_{\uparrow,j\pm1}$ be the speed of the gust front at the collision point, emitted by the $j\pm1$ convection. It represents the mechanical lifting effect. Similarly, we let $q'^n_{\uparrow,j\pm1}$ be the horizontal anomaly of the water vapor mixing ratio of the two involved gust fronts at the collision point, representing the thermodynamic forcing effect. Apart from the cold pool, we let $u_*$ and $q_*$ be the characteristic convergence wind speed and water vapor mixing ratio anomaly randomly produced by convective cells. The sub-cloud moisture convergence is expressed as:
\begin{equation}   \label{eq:moisture_convergence}
    Q^{n+1}_{\uparrow,j} = \underbrace{ \rho L_v \tau_\uparrow S }_{\text{constant}} \,\, \times \,\, 
\frac{1}{2} \left[ \left( u^n_{\uparrow,j-1} + u_* \right) \left( q'^n_{\uparrow,j-1} + q_* \right) + \left( u^n_{\uparrow,j+1} + u_* \right) \left( q'^n_{\uparrow,j+1} + q_* \right) \right]
\end{equation}
Here, $\rho$ is the air density, and $L_v$ is the vaporization heat. The $\tau_\uparrow$ is the duration of sub-cloud moisture convergence. Its upper bound is limited by the downdraft onset time, the time needed to generate a new cold pool and replace the old cold pools. The $S$ is the characteristic cross-sectional area for the convergent flow to enter the updraft.\footnote{Suppose the convergent wind transports air into a cylinder laterally and then upward into an updraft. The cylinder takes the updraft radius and the mixed layer thickness. The $S$ denotes the cylinder's lateral area.} The $\rho$, $L_v$, $\tau_\uparrow$, and $S$ are assumed to be constant quantities. Next, we study what determines $u^n_{\uparrow,j\pm1}$ and $q'^n_{\uparrow,j\pm1}$.

\subsubsection{The momentum balance of a gust front}
Because the momentum balance of a gust front is primarily between inertial force and pressure gradient force, $u^n_{\uparrow,j\pm1}$ depends on the horizontal anomaly of potential temperature at the gust front $\theta'^n_{\uparrow,j\pm1}$ \cite[e.g.,][]{benjamin1968gravity,ross2004theory,torri2015dyna_thermodyna}:
\begin{equation}  \label{eq:ustar}
    u^n_{\uparrow,j\pm1} = \mathrm{Fr} \left(-g\frac{\theta'^n_{\uparrow,j\pm1}}{\overline{\theta}} H_f \right)^{1/2}.
\end{equation}
Here, Fr represents the Froude number which is nearly unity for a gust front in tropical convection \citep{torri2019cold}, $g=9.8$ m s$^{-2}$ is the gravitation acceleration constant, $\overline{\theta}$ is the horizontally averaged $\theta$ in the mixed layer, and $H_f$ is the thickness of the gust front which is about half the mixed layer depth \citep{emanuel1994book}. We let Fr, $g$, $\overline{\theta}$, and $H_f$ be constant quantities. Thus, modeling $u^n_{\uparrow,j\pm1}$ turns to modeling $\theta'^n_{\uparrow,j\pm1}$. Next, we model $\theta'^n_{\uparrow,j\pm1}$ and $q'^n_{\uparrow,j\pm1}$ and show that they are closely related.

\begin{figure}[h] \centerline{\includegraphics[width=0.7\linewidth]{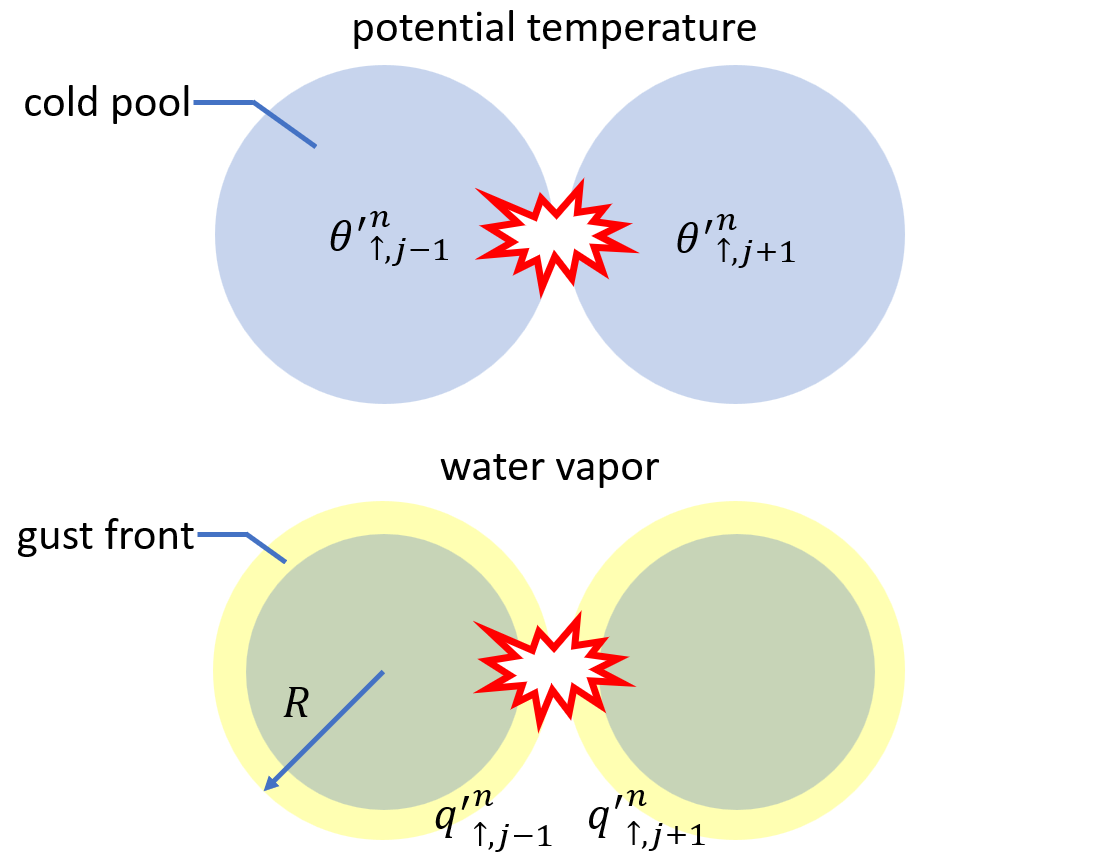}}
\caption{A schematic diagram of the potential temperature and water vapor field of two colliding cold pools with a radius of $R$, in a bird-eye view. The horizontal anomaly of potential temperature is assumed to be homogeneous in the cold pool, taking $\theta'^n_{\uparrow,j\pm1}$. The water vapor is anomalously high at the gust front, denoted as $q'^n_{\uparrow,j\pm1}$. }   \label{fig:cartoon_ring}
\end{figure}

\subsubsection{The bulk model for cold pool potential temperature}

The bulk model of \citet{romps2016sizes} assumes the cold pool as a cylinder with uniform buoyancy, as illustrated in Fig. \ref{fig:cartoon_ring}. The cylinder collapses and widens, gets diluted by entrainment, and gets dissipated by wind-induced surface heating, as depicted by their equation (27):  
\begin{equation}   \label{eq:theta_solution}
\begin{split}
    \theta'^n_{\uparrow,j\pm1} 
    &\approx 
    \underbrace{  \theta'^n_{\downarrow,j\pm1} \, e^{-\varepsilon {R}}  }_{\text{rain evap cooling}}
    + 
    \underbrace{ \frac{2\pi}{9}\frac{{C} \Delta \theta}{{V}} {R}^3\,e^{-\varepsilon {R}}\,\,  }_{\text{surface heating}}.
\end{split}    
\end{equation}
The initial cold pool radius factor is neglected because it is much smaller than $R$. Here $\theta'^n_{\downarrow,j\pm1}$ (negative) is the initial potential temperature anomaly of the cold pool, which is produced by rain evaporative cooling and is related to $Q^n_{\downarrow,j\pm1}$ (positive) through:
\begin{equation}   \label{eq:Qe_theta}
    Q^n_{\downarrow,j\pm1} = - \rho V c_p  \theta'^n_{\downarrow,j\pm1}.
\end{equation}
The parameter $c_p$ is the isobaric specific heat of air, ${C}$ is the nondimensional surface heat exchange coefficient, $\Delta \theta$ is the characteristic potential temperature difference between the cold pool and the sea surface, ${V}$ is the initial volume of the cold pool, and $\varepsilon$ (unit: m$^{-1}$) is the cold pool fractional entrainment rate which \citet{romps2016sizes} diagnosed to be a relatively fixed value of $\varepsilon\approx0.2$ km$^{-1}$. The parameters $\rho$, $c_p$, ${C}$, $\Delta \theta$, ${V}$, and $\varepsilon$ are assumed to be constant quantities.\footnote{The constant assumption of $\Delta \theta$, which neglects the anomalous air-sea temperature difference in and out of cold pools, has been used in the cold pool model of \citet{romps2016sizes}. Figure S2 in the supplemental dataset confirms that $\Delta \theta$ varies less than 50\% in and out of cold pools in example CM1 simulations ($\mathrm{E_v}=0.2$, 1, and 2.5) at $t=2.5$ days, so this is only a crude approximation.}

\subsubsection{A modified bulk model for gust front humidity}

The evolution of water vapor at the gust front (cold pool edge) is similar yet different from potential temperature. As shown in Figs. \ref{fig:pcolor} and \ref{fig:cartoon_ring}, the cold pool edge has a higher water vapor mixing ratio than the interior. Both the rain evaporation and surface evaporation contribute to this water vapor ring. Rain evaporation primarily humidifies the cold pool air initially in the mixed layer, which constitutes the gust front \citep{langhans2015ring}. Meanwhile, the water evaporated from the surface beneath the cold pool is preferentially transported to the gust front by the cold pool's internal circulation \citep{yao1996experimental,langhans2015ring}. The simplest treatment might be to assume these two effects balance out and directly apply the bulk model to the gust front quantity:
\begin{equation}   \label{eq:q_solution}
\begin{split}
    q'^n_{\uparrow,j\pm1} 
    &\approx 
\underbrace{q'^n_{\downarrow,j\pm1} \, e^{-\varepsilon {R}} }_{\text{rain evaporation}}
    + 
    \underbrace{ \frac{2\pi}{9}\frac{{C} \Delta q}{{V}} {R}^3 \,e^{-\varepsilon {R}} }_{\text{surface evaporation}},
\end{split}    
\end{equation}
where $\Delta q$ is the characteristic water vapor mixing ratio difference between the cold pool and the saturated air just above the sea surface, taken as a constant here. Here, $q'^n_{\downarrow,j\pm1}$ (positive) is the water vapor mixing ratio anomaly upon cold pool formation, and $q'^n_{\uparrow,j\pm1}$ is that upon cold pool collision. Note that they differ from the water vapor mixing ratio at the cold-pool interior, which shows a negative anomaly. The $q'^n_{\downarrow,j\pm1}$ is related to $\theta'^n_{\downarrow,j\pm1}$ via the approximate conservation of moist enthalpy during rain evaporation:
\begin{equation}  \label{eq:enthalpy}
    Q^n_{\downarrow,j\pm1} = - \rho V c_p  \theta'^n_{\downarrow,j\pm1}
    = \rho V L_v 
    q'^n_{\downarrow,j\pm1}.
\end{equation}
We have assumed that the water vapor mixing ratio and potential temperature of the mixed-layer air, before being influenced by sub-cloud evaporation, have relaxed to the horizontally averaged values. The first term of Eq. (\ref{eq:q_solution}) denotes rain evaporative moistening at the cold pool formation stage, which carries ``memory" from the previous lifecycle. The second term of Eq. (\ref{eq:q_solution}) denotes surface evaporative moistening, an approximately constant forcing term. Though \citet{langhans2015ring} used large-eddy simulations (LES) to show that the gust front moisture is contributed more by surface evaporative moistening than rain evaporative moistening, we retain both terms and study how they contribute to cloud interaction.

In summary, $\theta'^n_{\uparrow,j\pm1}$ and $q'^n_{\uparrow,j\pm1}$ are contributed by an initial component produced by rain evaporation and a forced component associated with wind-induced surface flux. The initial component is viewed as a variable that carries memory from the previous lifecycle. The accumulated surface flux is fixed because the cold pool collision radius ${R}$ is fixed. For potential temperature, the initial and forced components have the opposite sign. For the water vapor mixing ratio, the sign is the same. This difference will be shown to be crucial in determining the nonlocal influence length scale of the air column humidity. Model parameters are summarized in Table \ref{Table_parameters}.

\begin{table}[h]
\caption{A summary of parameters in the cloud chain model. }\label{Table_parameters}
\begin{center}
\begin{tabular}{ccccrrcrc}
\hline
\hline
Name  &  Meaning & Appearance & Value \\
\hline
$\eta_*$ & Equilibrium ratio of sub-cloud evaporation to moisture convergence & Eq. (\ref{eq:eta}) & $0<\eta_*<1$ \\ 
$\mathrm{PW}_*$ & The PW measuring the sensitivity of sub-cloud evaporation to moisture convergence & Eq. (\ref{eq:eta}) & $\sim 1$ mm \\
$S$ & Characteristic cross-sectional area for convergent to enter the updraft & Eq. (\ref{eq:moisture_convergence}) & $\sim$1 km$^2$ \\
$\tau_{\uparrow}$ & Duration time of sub-cloud moisture convergence & Eq. (\ref{eq:moisture_convergence}) & $\sim$ 0.5 hours \\
$\rho$  & Mixed-layer air density & Eq. (\ref{eq:moisture_convergence}) & $\sim 1$ kg m$^{-3}$ \\
$L_v$  & Vaporization heat & Eq. (\ref{eq:moisture_convergence}) & $2.5 \times 10^6$ J kg$^{-1}$ \\
$u_*$  & Characteristic convergence wind speed of mixed layer convective cell & Eq. (\ref{eq:moisture_convergence}) & $\sim$1 m s$^{-1}$ \\
$q_*$ & Characteristic water vapor mixing ratio anomaly at the convergent region & Eq. (\ref{eq:moisture_convergence}) & $\sim1$ g kg$^{-1}$ \\
$g$ & Gravitational acceleration constant & Eq. (\ref{eq:ustar}) & 9.8 m s$^{-2}$ \\
$\overline{\theta}$ & Horizontally averaged potential temperature in the mixed layer & Eq. (\ref{eq:ustar}) & $\sim$300 K\\
$H_f$ & Thickness of a gust front & Eq. (\ref{eq:ustar}) & $\sim 0.25$ km  \\
Fr  & Froude number of a cold pool & Eq. (\ref{eq:ustar}) & $\sim$ 1 \\
$\varepsilon$  & Fractional entrainment rate of a cold pool & Eq. (\ref{eq:theta_solution}) & $\sim0.2$ km$^{-1}$ \\
$C$ & Surface heat exchange coefficient & Eq. (\ref{eq:theta_solution}) & $\sim 0.001$ \\
$\Delta \theta$ & Air-sea surface temperature difference & Eq. (\ref{eq:theta_solution}) & $\sim 1$ K \\
$c_p$  & Isobaric specific heat & Eq. (\ref{eq:Qe_theta}) & $1005$ J kg$^{-1}$ K$^{-1}$ \\
$V$  & Initial cold pool volume & Eq. (\ref{eq:Qe_theta}) & $\sim1$ km$^3$ \\
$\Delta q$ & Air-sea water vapor mixing ratio difference & Eq. (\ref{eq:q_solution}) & $\sim1$ g kg$^{-1}$ \\
$\alpha$ & Fractional contribution of cold pools to convective initiation & Eq. (\ref{eq:alpha}) & $0<\alpha<1$ \\
\hline
\hline
\end{tabular}
\end{center}
\end{table}

\subsection{The cloud chain model}

With models of the convective and cold-pool phases at hand, we can combine them into a cloud-chain model and then use a perturbation analysis to study the nonlocal influence length scale. 

\subsubsection{The fully nonlinear form}

Equations (\ref{eq:moisture_convergence})-(\ref{eq:enthalpy}) can be combined into a single chain equation of the sub-cloud rain evaporation amount $Q^n_{\downarrow,j}$, which represents the initial intensity of each round of cold pool. Three steps of preparation are needed.

First, all variables in the cloud chain model are decomposed into an equilibrium component ($\langle ... \rangle$) and a perturbation component ($\delta...$):
\begin{equation}  \label{eq:equilibrium_decompose}
\begin{split}
    q'^n_{\downarrow,j} &\equiv \langle q'_{\downarrow} \rangle + \delta q'^n_{\downarrow,j}, \quad q'^n_{\uparrow,j} \equiv \langle q'_{\uparrow} \rangle + \delta q'^n_{\uparrow,j},\\
    \theta'^n_{\downarrow,j} &\equiv \langle \theta'_{\downarrow} \rangle + \delta \theta'^n_{\downarrow,j},\quad \theta'^n_{\uparrow,j} \equiv \langle \theta'_{\uparrow} \rangle + \delta \theta'^n_{\uparrow,j},\\
    u^n_{\downarrow,j} &\equiv \langle u_{\downarrow} \rangle + \delta u^n_{\downarrow,j}, \quad u^n_{\uparrow,j} \equiv \langle u_{\uparrow} \rangle + \delta u^n_{\uparrow,j},\\
    Q^n_{\downarrow,j} &\equiv \langle Q_{\downarrow} \rangle + \delta Q^n_{\downarrow,j}, \,\,\, Q^n_{\uparrow,j} \equiv \langle Q_{\uparrow} \rangle + \delta Q^n_{\uparrow,j}.    
\end{split}    
\end{equation}
The equilibrium state has homogeneous, steady convective activity, with quantities identical across different $j$ and $n$. Note that the perturbation component is not assumed to be small.

Second, we define $F_{\theta,\mathrm{rain}}$ and $F_{\theta,\mathrm{surf}}$ (unit: J kg$^{-1}$) as the total rain evaporative cooling and total surface sensible heating experienced by per unit mass of a cold pool. Similarly, we define $F_{q,\mathrm{rain}}$ and $F_{q,\mathrm{surf}}$ as the total rain evaporation and total surface evaporation experienced by per unit mass of a cold pool, expressed in the form of latent heat (multiplied with $L_v$):
\begin{equation}  \label{eq:rain_surf}
\begin{split}
F_{\theta,\mathrm{rain}} &\equiv L_v \langle{q'_{\downarrow}}\rangle, \\
F_{\theta,\mathrm{surf}} &\equiv c_p \frac{2\pi}{9}\frac{{C} \Delta \theta}{{V}}{R}^3, \\   F_{q,\mathrm{rain}} &\equiv L_v \langle{q'_{\downarrow}}\rangle, \\
F_{q,\mathrm{surf}} &\equiv L_v \frac{2\pi}{9}\frac{{C} \Delta q}{{V}}{R}^3.
\end{split}    
\end{equation}
In expressing the rain evaporative cooling term $F_{\theta,\mathrm{rain}}$, the enthalpy conservation relation in rain evaporation [Eq. (\ref{eq:enthalpy})] has been used. We further define the ratio of rain evaporative cooling to surface sensible heating as $\mathcal{R}_{\theta}$ and similarly define the ratio of rain evaporation to surface evaporation as $\mathcal{R}_{q}$:
\begin{equation}
    \mathcal{R}_{\theta} \equiv \frac{F_{\theta,\mathrm{rain}}}{F_{\theta,\mathrm{surf}}},
    \quad
    \mathcal{R}_{q} \equiv \frac{F_{q,\mathrm{rain}}}{F_{q,\mathrm{surf}}}.
\end{equation}
Using [Eq. (\ref{eq:enthalpy})] again, we find that $\mathcal{R}_{\theta}/\mathcal{R}_{q}$ is essentially the ratio of surface latent heat flux to sensible heat flux, i.e., the inverse of the Bowen ratio $B$:
\begin{equation}  \label{eq:Bowen}
    \frac{\mathcal{R}_{\theta}}{\mathcal{R}_{q}}
    = \frac{F_{q,\mathrm{surf}}}{F_{\theta,\mathrm{surf}}}
    = B^{-1}.
\end{equation}

Third, the parameter $\alpha$ is introduced as the fractional contribution of cold pools to convective initiation:
\begin{equation}  \label{eq:alpha}
    \alpha = \frac{\langle u_{\uparrow} \rangle }{\langle u_{\uparrow} \rangle + u_*}
    = \frac{\langle q'_{\uparrow} \rangle }{\langle q'_{\uparrow} \rangle + q_*},
\end{equation}
based on the assumption that the fractional contribution of cold pools to mechanical lifting and thermodynamic forcing is identical. Suppose the cold pool is too negatively buoyant upon collision, as is the case in some mesoscale convective systems. In that case, the air parcel involved in the sub-cloud moisture convergence will no longer be the gust front air, but the mixed-layer air beyond the front instead \citep{droegemeier1985three,tompkins2001organization}. This would reduce the fractional contribution of cold pools to thermodynamic forcing and, in turn, reduce $\mathcal{W}$. This regime is left for future study. Based on Fig. \ref{fig:theta_KE}, we express $\alpha$ as the deviation from the gust front momentum balance [Eq. (\ref{eq:ustar})]:
\begin{equation}  \label{eq:beta_empirical}
    \alpha \approx 2.5 \frac{\mathrm{std}(\theta_s)}{\overline{\mathrm{KE}_s}}.
\end{equation}
This data-driven expression will be used in section \ref{sec:validation} for model validation. The 2.5 factor is an empirical factor that calibrates $\alpha$ to be close to $1$ in the $\mathrm{E_v}\gg 1$ (very-strong-cold-pool) regime.

Substituting Eqs. (\ref{eq:eta}) and (\ref{eq:ustar})-(\ref{eq:alpha}) into Eq. (\ref{eq:moisture_convergence}), without any approximations, we obtain a simple form of the cloud chain, where the perturbation component of the sub-cloud rain evaporation amount $\delta Q^{n}_{\downarrow,j}$ is the prognostic variable:
\begin{equation}  
\label{eq:nondim_chain}
\begin{split}
     \frac{\delta Q^{n+1}_{\downarrow,j}}{\langle Q_{\downarrow} \rangle} &= \left( 1 + \frac{\mathrm{PW}'_j}{\mathrm{PW}_*} \right) \\
     &\times \frac{1}{2}\sum_{k=-1,1} \left\{ \underbrace{ \left\{ 1 + \alpha \left[ \left( 1 + \frac{1}{1-1/\mathcal{R}_\theta} \frac{\delta Q^n_{\downarrow,j+k}}{\langle Q_\downarrow \rangle} \right)^{1/2} - 1 \right] \right\}}_{\text{mech lifting}} \underbrace{\left\{ 1 + \frac{\alpha}{1+1/\mathcal{R}_q} \frac{\delta Q^n_{\downarrow,j+k}}{\langle Q_\downarrow \rangle} \right\}}_{\text{thermo forcing}} - 1 \right\}.
\end{split}    
\end{equation}
See Appendix C for derivation details. The nondimensional quantity $\delta Q^{n+1}_{\downarrow,j}/\langle Q_{\downarrow} \rangle$ is sufficient for understanding the model behavior.

To understand the model behavior, we perform a numerical simulation of the cloud chain model, which serves as a quantitative illustration of the cartoon shown in Fig. \ref{fig:cartoon_spread}. The chain has $M=20$ cloud members, $j=1$, 2, ..., $M$. They are linked periodically. The cold pool related parameters use $\mathcal{R}_q=0.4$, $B=0.1$ (equivalently, $\mathcal{R}_{\theta}=4$), and $\alpha=1$. In section \ref{sec:validation}, they will be shown relevant to the $\mathrm{E_v}=1$ reference cloud-permitting simulation. The forcing is prescribed as 
\begin{equation}
    \frac{\mathrm{PW'}_j}{\mathrm{PW}_*}=
    \begin{cases}
        \frac{1}{4},\quad j = \frac{N}{2},\\
        0, \quad \mathrm{else}.
    \end{cases}
\end{equation}
The initial condition is the equilibrium state of the case without forcing: 
\begin{equation}
    \frac{\delta Q^{n=1}_{\downarrow,j}}{\langle Q_{\downarrow} \rangle} = 
    \begin{cases}
        0, \quad j\in\text{even}, \\
        \text{not defined}, \quad j\in\text{odd}. 
    \end{cases}
\end{equation}
Half of the grid points are not defined because convective evolution depends on the collision of previous neighbors and is therefore staggered in space and time. 

\begin{figure}[h]
\centerline{\includegraphics[width=0.55\linewidth]{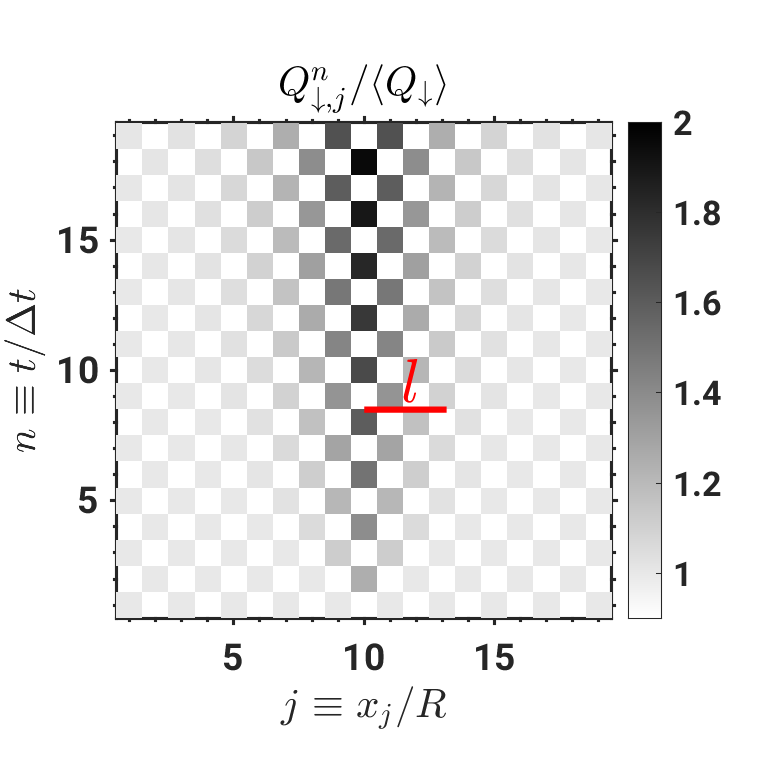}}
  \caption{An example numerical simulation of the fully nonlinear cloud chain model [Eq. (\ref{eq:nondim_chain})]. The horizontal axis shows the cloud position, and the vertical axis shows the lifecycle number. The checkerboard shows the sub-cloud rain evaporation amount in each cold pool $Q^n_{\downarrow,j\pm1}$, rescaled by the equilibrium component $\langle Q_{\downarrow} \rangle$. It includes both the equilibrium and perturbation components. The white regions denote the spatiotemporal positions where $Q^n_{\downarrow,j}$ is not defined. The red horizontal line illustrates the nonlocal influence length scale $l$.} \label{fig:diff_simulation}
\end{figure}

Figure \ref{fig:diff_simulation} shows the $Q^{n}_{\downarrow,j}$ (equilibrium plus perturbation) of the first 20 convective lifecycles in the simulations. The cloud chain reaches a relatively steady state within 10 cycles. The perturbation is most substantial at the forcing location, but it laterally spreads over a few cloud spacings. The evolution appears to be governed by a combination of diffusive and damping processes. Next, we perform a small perturbation analysis to further quantify the cloud chain's behavior.

\subsection{The small perturbation analysis}\label{subsec:small_perturbation}

Assuming $\delta Q^n_{\downarrow,j} / \langle Q_{\downarrow} \rangle \ll 1$ and $\mathrm{PW'}_j / \mathrm{PW}_* \ll 1$, we approximate the fully nonlinear cloud chain model [Eq. (\ref{eq:nondim_chain})] as a linear chain model:
\begin{equation}  \label{eq:linear_chain_down}
\begin{split}
    \frac{\delta Q^{n+1}_{\downarrow,j}}{\langle Q_{\downarrow} \rangle}
&\approx \frac{1}{2} 
    \sum_{k=-1,1} 
 \left( 1 + \frac{\delta Q^{n}_{\downarrow,j+k}}{\langle Q_{\downarrow} \rangle} \frac{\alpha}{2}\frac{1}{1-B/\mathcal{R}_q} + \frac{\delta Q^{n}_{\downarrow,j+k}}{\langle Q_{\downarrow} \rangle} \frac{\alpha}{1+1/\mathcal{R}_q} + \frac{\mathrm{PW'}_j}{\mathrm{PW}_*} \right) - 1 \\
 &= 
    \frac{\mathcal{W}}{2} \left( \frac{\delta Q^n_{\downarrow,j-1}}{\langle Q_{\downarrow}\rangle} + \frac{\delta Q^n_{\downarrow,j+1}}{\langle Q_{\downarrow}\rangle} \right) 
    + \frac{\mathrm{PW'}_j}{\mathrm{PW}_*}.
\end{split}
\end{equation}
Here, $\mathcal{W}$ is referred to as the convective memory weight, which consists of the mechanical lifting part $\mathcal{W}_{\theta}$ and the thermodynamic forcing part $\mathcal{W}_q$:
\begin{equation}   \label{eq:chain_weight}
\begin{split}
    \mathcal{W} 
    &\equiv \alpha  \left( \underbrace{  \frac{1}{2}\frac{1}{1-B/\mathcal{R}_q}  }_{\text{mech lifting},\,\,\mathcal{W}_{\theta}}
    \,\,+ \,\,
    \underbrace{ \frac{1}{ 1+1 / \mathcal{R}_q } }_{\text{thermo forcing},\,\,\mathcal{W}_{q}} \right).
\end{split}    
\end{equation}
The cold pool contribution fraction $\alpha$ serves as a multiplication factor to $\mathcal{W}_{\theta}+\mathcal{W}_{q}$. Equation (\ref{eq:linear_chain_down}) shows how a locally humid air column ($\mathrm{PW}'_j$) influences distant sub-cloud evaporation. The chain can be rephrased as a finite-difference equation:
\begin{equation}   \label{eq:FD}
        \frac{\delta Q^{n+1}_{\downarrow,j} - \delta Q^{n}_{\downarrow,j}}{\Delta t} = \frac{\mathcal{W} R^2}{2 \Delta t} \frac{ \delta Q^{n}_{\downarrow,j-1} - 2 \delta Q^{n}_{\downarrow,j} + \delta Q^{n}_{\downarrow,j+1} }{R^2}
        - \frac{1-\mathcal{W}}{\Delta t} \delta Q^{n}_{\downarrow,j}
        +\frac{\langle Q_{\downarrow} \rangle}{\Delta t} \frac{\mathrm{PW'}_j}{\mathrm{PW}_*}.
\end{equation}
If we assume that the system evolves slowly, so that $\delta Q^{n+1}_{\downarrow,j} \approx \delta Q^{n}_{\downarrow,j}$, and view $\delta Q^{n}_{\downarrow,j}$ and $\mathrm{PW'}_j$ as continuous functions in space ($\delta Q_{\downarrow}$ and $\mathrm{PW'}$), Eq. (\ref{eq:FD}) reduces to a damping-diffusion equation:
\begin{equation}  \label{eq:reaction_diffusion_1D}
    -l^2 \frac{\partial^2}{\partial x^2} \delta Q_{\downarrow} + \delta Q_{\downarrow} = \delta Q^\#_{\downarrow}.
\end{equation}
Here, $\delta Q^\#_{\downarrow}$ is the $\delta Q_{\downarrow}$ when its horizontal gradient vanishes:
\begin{equation}
    \delta Q^\#_{\downarrow} = \frac{\langle Q_{\downarrow} \rangle}{1-\mathcal{W}} \frac{\mathrm{PW'}}{\mathrm{PW}_*}.
\end{equation}
The $\delta Q^\#_{\downarrow}$ represents the equilibrium response of sub-cloud rain evaporation amount to the air-column humidity anomaly, in the absence of a down-gradient diffusion.\footnote{We remind readers that $\delta Q^\#_{\downarrow}$ does \textit{not} represent the case where cold pools vanish.} A more humid air column fosters heavier precipitation and stronger cold pools. The $1/(1-\mathcal{W})$ is a ``gain factor" that represents the memory effect of previous convective events, with a higher $\mathcal{W}$ increasing $\delta Q^\#_{\downarrow}$. The diffusion effect is quantified as the nonlocal influence range $l$ of a humid air column on the convective/cold pool intensity:
\begin{equation}  \label{eq:Lcp_theory}
    {l}
    = \left( \frac{1}{2} \frac{\mathcal{W}}{1-\mathcal{W}} \right)^{1/2} {R},
\end{equation}
which only depends on the convective memory weight $\mathcal{W}$ and the mean cold pool radius ${R}$.

\subsection{The convective memory weight}

\begin{figure}[h]
 \centerline{\includegraphics[width=1\linewidth]{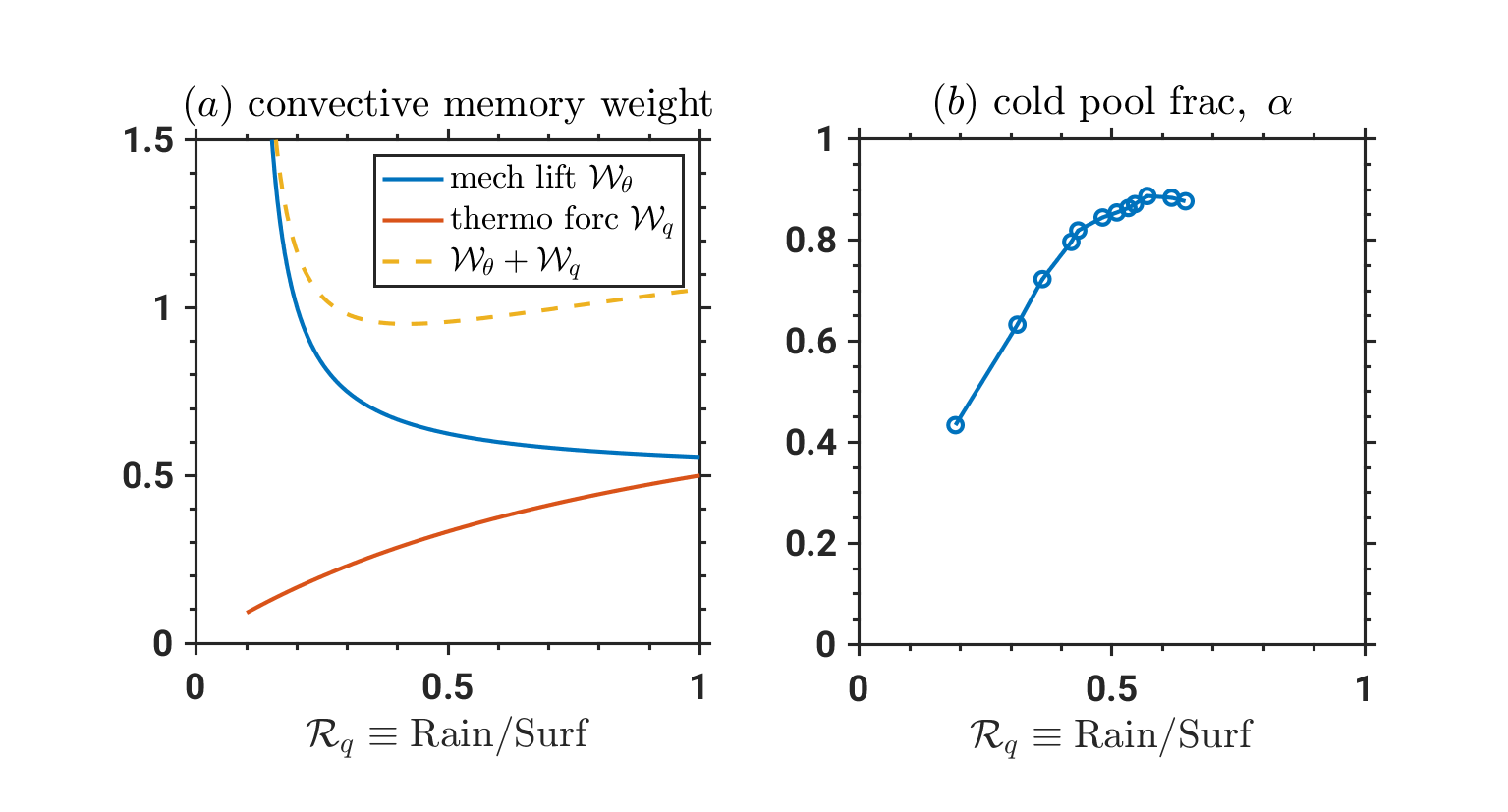}}
  \caption{(a) Theoretical calculations of the convective memory weight, using $B=0.1$. The solid blue line shows the mechanical lifting part $\mathcal{W}_{\theta}$, the solid red line shows the thermodynamic forcing part $\mathcal{W}_q$, and the dashed yellow line shows their sum. (b) The fractional contribution of cold pools to convective initiation, $\alpha$, versus $\mathcal{R}_q$, diagnosed with Eq. (\ref{eq:beta_empirical}). } \label{fig:cartoon_hook}
\end{figure}

So far, $\mathcal{W}$ has been expressed as a function of three nondimensional parameters: $\mathcal{R}_q$, $B$, and $\alpha$. For fixed $B$, the mechanical lifting effect $\mathcal{W}_{\theta}$ decreases with $\mathcal{R}_q$, and the thermodynamic forcing effect $\mathcal{W}_{q}$ increases with $\mathcal{R}_q$ (Fig. \ref{fig:cartoon_hook}a). Their sum has a hook-shaped dependence on $\mathcal{R}_q$. The differing dependence trend of $\mathcal{W}_{\theta}$ and $\mathcal{W}_{q}$ on $\mathcal{R}_q$ is due to the fundamentally different effects of rain evaporation and surface flux on potential temperature and humidity. For potential temperature, rain evaporative cooling and surface sensible heating have opposite effects. For humidity, rain evaporation and surface evaporation have aligned effects. More explanations are provided below, supposing fixed $B$.\footnote{The constant $B$ assumption here is for theoretical understanding only. Section \ref{sec:validation} will show that $B$ slightly increases with $\mathrm{E_v}$ in cloud-permitting simulations.}  

The trend of $\mathcal{W}_{\theta}+\mathcal{W}_q$ in the $\mathcal{R}_q\lesssim 0.4$ regime is dominated by mechanical lifting, and that in the $\mathcal{R}_q\gtrsim 0.4$ regime is dominated by thermodynamic forcing. 
\begin{itemize}
    \item For the mechanical lifting effect, an \textit{overall} stronger surface sensible heating (smaller $\mathcal{R}_{\theta}$ or $\mathcal{R}_q$) makes cold pools less negatively buoyant when colliding. In other words, convective initiation gets more sensitive to the collision speed. A perturbation in the initial potential temperature of a \textit{specific} cold pool, which directly induces a perturbation in the collision speed, will more significantly influence convective initiation. Thus, a smaller $\mathcal{R}_{q}$ \textit{enhances} the mechanical lifting component of the convective memory weight $\mathcal{W}_{\theta}$.
    \item For the thermodynamic forcing effect, an \textit{overall} stronger surface evaporation (smaller $\mathcal{R}_q$) makes cold pools more humid when colliding. In other words, convective initiation becomes less sensitive to humidity perturbation at collisions. A perturbation in the initial humidity of a \textit{specific} gust front, which directly induces a perturbation in the collisional humidity, will less significantly influence convective initiation. Thus, a smaller $\mathcal{R}_{q}$ \textit{reduces} the thermodynamic forcing component of the convective memory weight $\mathcal{W}_{q}$.
\end{itemize}

The non-monotonic dependence of $\mathcal{W}_{\theta}+\mathcal{W}_q$ on $\mathcal{R}_q$ does not necessarily mean that $\mathcal{W}$ has a non-monotonic dependence on $\mathcal{R}_q$, because $\alpha$ increases significantly with rain evaporation. A greater $\mathrm{E_v}$ indicates greater $\mathcal{R}_q$, making the moisture convergence in the mixed layer depend more on cold pools and increasing $\alpha$ (Fig. \ref{fig:cartoon_hook}b). The relative importance of the two factors need to be evaluated with data. 

%The monotonic increase of $\alpha$ with $\mathcal{R}_q$, combined with the non-monotonic change of $\mathcal{W}_{\theta}+\mathcal{W}_q$ with $\mathcal{R}_q$, determine the trend of $\mathcal{W}$. 

\subsection{A data-driven theoretical inference of $l$}\label{subsec:diagnose_memory}

The nonlocal influence length scale of a humid air column $l$ depends on $\mathcal{W}$ and $R$ [Eq. (\ref{eq:Lcp_theory})], and $\mathcal{W}$ depends on $\mathcal{R}_{q}$, $\mathcal{R}_{\theta}$, and $\alpha$ [Eq. (\ref{eq:chain_weight})]. The strategy is to perform a data-driven theoretical inference of $\mathcal{W}$ using cloud-permitting simulations.

\begin{figure}[h]
\centerline{\includegraphics[width=1.3\linewidth]{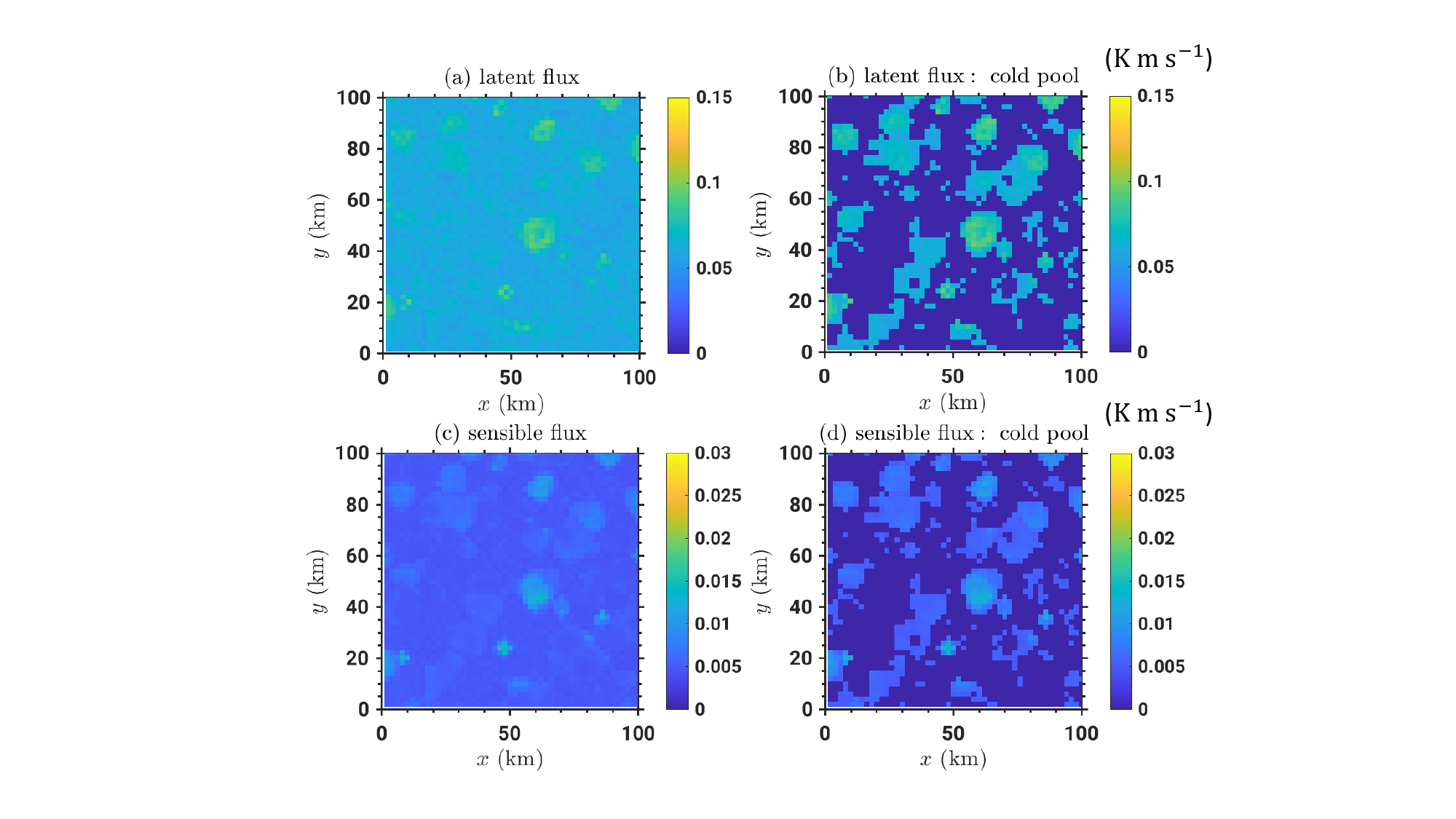}}
  \caption{Zoom-in plots of surface heat flux variables in the 100 km $\times$ 100 km southwest corner region, using the $\mathrm{E_v}=1$ reference experiment at $t=2.5$ days. All variables have been divided by $c_p=1005$ J kg$^{-1}$ K$^{-1}$ and shown in the unit of K m s$^{-1}$. (a) The latent heat flux. (b) The latent heat flux in the cold pool region. (c) The sensible heat flux. (d) The sensible heat flux in the cold pool region. }\label{fig:flux}
\end{figure}   
% Fig. \ref{fig:cartoon_hook} shows that 

% The $XXX$, which must be larger than unity, is the most uncertain parameter. Using a typical Bowen ratio of $B=0.1$ for the tropical oceanic environment \cite[e.g.,][]{jo2002Bowen,fu2025mean_part1}, focusing on the physically relevant $\mathcal{R}_q > B$ regime (will be verified with simulations in section \ref{sec:validation}\ref{subsec:diagnose_memory}), fixing $\alpha=1/2$ as the strong cold pool regime, and trying $XXX=1$, 1.5, and 2, we find $\mathcal{W}$ obeys a hook shape (Fig. \ref{fig:cartoon_hook}):
% \begin{itemize}
%     \item For $\mathcal{R}_q\lesssim 0.4$, $\mathcal{W}$ drops as $\mathcal{R}_q$ increases; $\mathcal{W}_{\theta}$ dominates $\mathcal{W}$.
%     \item For $\mathcal{R}_q\gtrsim 0.4$, $\mathcal{W}$ rises as $\mathcal{R}_q$ increases; $\mathcal{W}_{q}$ dominates $\mathcal{W}$.
% \end{itemize}

We introduce the diagnostic method of $\mathcal{R}_{q}$ first, with a similar method for $\mathcal{R}_{\theta}$. The $\mathcal{R}_{q}$ depend on the ratio of rain evaporation amount per unit mass of a cold pool, $F_{q,\mathrm{rain}}$, to the surface evaporation amount per unit mass of a cold pool $F_{q,\mathrm{surf}}$. For $F_{q,\mathrm{rain}}$, we calculate the rain evaporative moistening rate [``$\mathrm{qt\_evar}$", unit: kg kg$^{-1}$ s$^{-1}$] and rain evaporative cooling rate [``$-\mathrm{tt\_evar}$", unit: K s$^{-1}$] in the lowest eight grid levels, i.e., at and below $z=0.955$ km. The choice of this height is based on the established understanding that the mean original height of cold pool air in tropical convection is around 1 km \citep{torri2016lagrangian}. Because rain evaporation is mainly associated with cold pools, we do not need to specifically pick out cold pool regions when calculating $F_{q,\mathrm{rain}}$. For $F_{q,\mathrm{surf}}$, the cold pool regions must be picked out since the surface evaporation outside cold pools is non-negligible. A simple way is to let the cold pool region be the place where the horizontal anomaly of near-surface potential temperature ($\theta'_s$) is negative. The $F_{q,\mathrm{surf}}$ is calculated as the conditional average of the surface evaporation rate ``$\mathrm{qvflux}$" (unit: kg kg$^{-1}$ m s$^{-1}$) in the cold pool region. The same method is used for calculating $F_{\theta,\mathrm{rain}}$ and $F_{\theta,\mathrm{surf}}$, using the rain evaporative cooling rate ``tt\_evar" and the surface sensible heat flux ``$\mathrm{thflux}$" (unit: K m s$^{-1}$). Figure \ref{fig:flux} shows that this simple method can crudely isolate the surface flux inside cold pool regions. The diagnostic method of $\mathcal{R}_q$ and $\mathcal{R}_{\theta}$ is summarized below:
\begin{equation}  \label{eq:ratio_diagnosis}
   \mathcal{R}_{q} = \frac{ \frac{1}{L^2}\int_0^L \int_0^L \int_0^{0.955\,\,\mathrm{km}} \, \mathrm{qt\_evar}\, dxdy}{\frac{1}{L^2}\int_0^L \int_0^L \, \mathrm{qvflux}\, \mathcal{H}(-\theta'_s)\,dxdy},
   \quad
   \mathcal{R}_{\theta} = \frac{ \frac{1}{L^2}\int_0^L \int_0^L \int_0^{0.955\,\,\mathrm{km}} \, -\mathrm{tt\_evar}\, dxdy}{\frac{1}{L^2}\int_0^L \int_0^L \, \mathrm{thflux}\, \mathcal{H}(-\theta'_s)\,dxdy},
\end{equation}
where $\mathcal{H}(-\theta'_s)$ is the Heaviside function that takes 1 when $\theta'_s<0$ and 0 when $\theta'_s>0$. Variation in air density within the lowest 0.955 km of the domain is ignored.

% Because the surface evaporation rate ``$\mathrm{qvflux}$" (unit: kg kg$^{-1}$ m s$^{-1}$) inside a cold pool is significantly higher than outside, and the evaporative flux outside takes a relatively uniform value (Fig. \ref{fig:flux}a)\footnote{The relatively uniform background surface flux is associated with the wind of convective cells in the mixed layer \citep{emanuel1994book}.}, a simple way for calculating evaporation in cold pool regions is to subtract the minimum surface evaporation rate in the domain from ``qvflux" (Fig. \ref{fig:flux}b), aware that the surface evaporation rate in cold pool regions is also subtracted a value. The surface sensible heat flux ``$\mathrm{thflux}$" (unit: K m s$^{-1}$) obeys a similar pattern to ``$\mathrm{qvflux}$", with anomalously high sensible heating inside cold pool regions and relatively uniform low-sensible heating outside. Thus, we use the same method for calculating $F_{\theta,\mathrm{rain}}$ and $F_{\theta,\mathrm{surf}}$. 

\begin{figure}[h] \centerline{\includegraphics[width=1.1\linewidth]{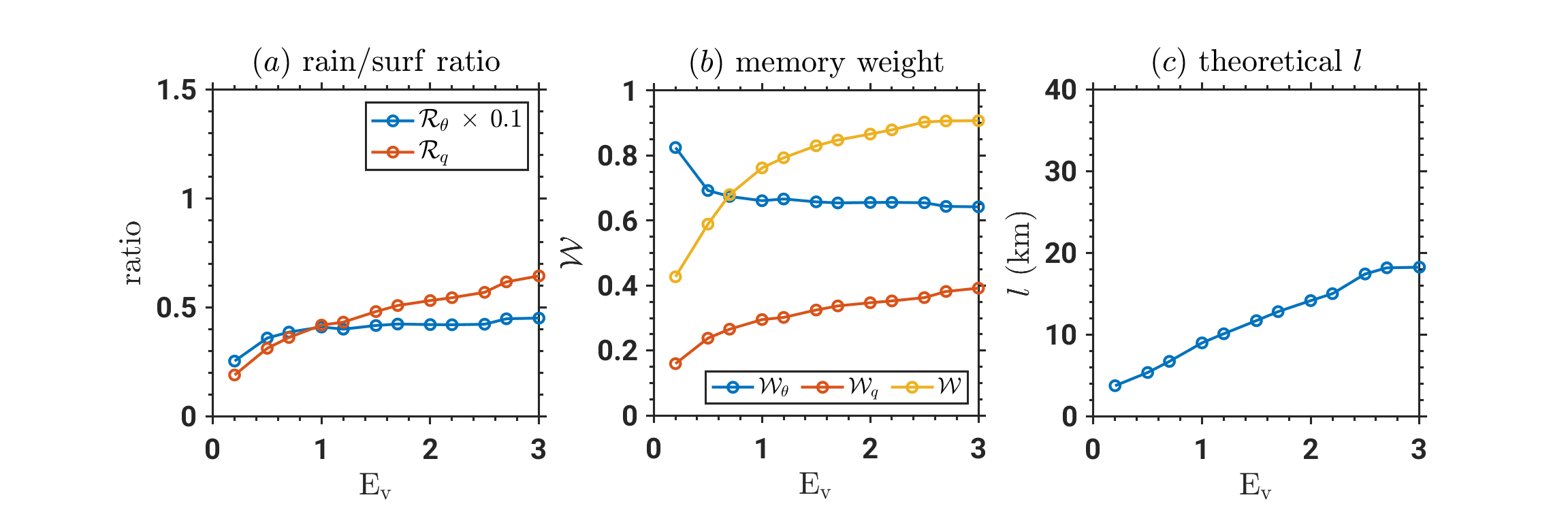}}
  \caption{(a) The ratio of rain evaporative cooling to surface sensible heating in cold pool regions $\mathcal{R}_{\theta}$ (blue line), and the ratio of rain evaporation to surface evaporation in cold pool regions $\mathcal{R}_q$ (red line), using the simulation with varying $\mathrm{E_v}$ at $t=2.5$ days. Note that $\mathcal{R}_{\theta}$ has been multiplied with a $0.1$ factor to facilitate comparison with $\mathcal{R}_{q}$. (b) The convective memory weight of the mechanical lifting component $\mathcal{W}_{\theta}$ (blue line), thermodynamic forcing component $\mathcal{W}_q$ (red line), and the total weight $\mathcal{W}$ (yellow line). They are calculated with Eq. (\ref{eq:chain_weight}), using the diagnosed $\mathcal{R}_{\theta}$ and $\mathcal{R}_q$ as shown in (a), and the diagnosed $\alpha$ from Eq. (\ref{eq:beta_empirical}). (c) The theoretically inferred $l$.} \label{fig:memory_weight}
\end{figure}

Figure \ref{fig:memory_weight}a shows the diagnostic result of $\mathcal{R}_q$ and $\mathcal{R}_{\theta}$ for experiments with varying $\mathrm{E_v}$. The ratio $\mathcal{R}_q$ increases from 0.2 to 0.65 as $\mathrm{E_v}$ increases, because stronger sub-cloud rain evaporation naturally increases the ratio of rain evaporation to surface evaporation in cold pool regions. The ratio $\mathcal{R}_{\theta}$ first increases from 2.5 to 4.5 and then stays around this value for $\mathrm{E_v}\gtrsim 1$. The $\mathcal{R}_{q}/\mathcal{R}_{\theta}$ is around 0.1, the same order of magnitude as the tropical marine Bowen ratio \citep{jo2002Bowen}. 

Figure \ref{fig:memory_weight}b shows the convective memory weight calculated with the diagnosed $\mathcal{R}_q$ and $\mathcal{R}_{\theta}$, using the diagnosed $\alpha$ [Eq. (\ref{eq:beta_empirical})]. As $\mathrm{E_v}$ increases, $\mathcal{W}_{\theta}$ decreases, $\mathcal{W}_{q}$ increases, and $\alpha$ increases (see Fig. \ref{fig:cartoon_hook}b). The net result is that $\mathcal{W}$ increases from around 0.4 to around 0.9. A greater $\mathcal{W}$ would make the cloud chain unstable. Readers may ask: What will happen if the rain evaporation rate further increases? The author has no answer now, but the Fourier spectral peak of the column precipitable water $|\hat{\mathrm{PW}}|$ indeed turns from decrease to slightly increase when $\mathrm{E_v} \gtrsim 2$ (Fig. \ref{fig:spectrum}b). This might be the case in which cold pool dynamics drive convective self-aggregation, as hypothesized by \citet{Haerter_2019}. To obtain a solid answer, the theory needs to consider the influence of convective inhibition energy \cite[CIN,][]{mapes2000toy,bretherton2004new,grandpeix2010density}. CIN may throttle the total water vapor transported to an updraft, $Q^n_{\uparrow,j}$. This may reduce the fraction of cold-pool air in the lifted air and, therefore, $\alpha$, reducing the convective memory weight.

The above theoretical inference of $l$ needs to be benchmarked with cloud-permitting simulations. Diagnosing $l$ directly is challenging. An ideal method is to isolate a single vortex and examine whether the gust front humidity is distributed more smoothly than $\mathrm{PW}'$. There are two complexities. First, diagnosing gust front (cold pool edge) humidity requires an algorithm to identify the cold pool edge. Second, TC embryo vortices form in a periodic pattern in the simulations, making the isolation analysis difficult. Meanwhile, the TC embryo size can be readily diagnosed from the Fourier spectrum (Fig. \ref{fig:spectrum}). Therefore, we proceed to derive the TC embryo size theory by coupling the cloud chain model and the QG model. Then, the TC embryo size theory is benchmarked with cloud-permitting simulations.

\section{The TC embryo vortex size}\label{sec:validation}

This section couples the cloud chain model with the QG model, performs a linear stability analysis, and benchmarks the predicted vortex size with cloud-permitting simulations. 

% The theoretical model is validated by comparing the theoretical cloud interacting length scale ${l}$ [Eq. (\ref{eq:Lcp_theory})] with the diagnostic result of CM1 cloud-permitting simulations (Fig. \ref{fig:spectrum}). There are three gaps to address before performing the validation.

%The condensation depends on both the sub-cloud moisture convergence and the air-column humidity \citep{derbyshire2004RH}. 

\subsection{Bridging the cold pool activity and diabatic heating}

The cloud chain model depicts the sub-cloud evaporation amount per cold pool. The horizontal anomaly of diabatic heating rate $J'_\mathrm{dia}$ depends on the rain formation amount per convective event. As discussed in section \ref{sec:theory}\ref{subsec:framework}, in a low-CAPE environment like the tropics, the amount of sub-cloud rain evaporation largely reflects the rain formation amount and, therefore, the convective diabatic heating amount \citep{james2010numerical}. In other words, convective and cold-pool activity are aligned. Thus, $J'_\mathrm{dia}$ is closely related to $\delta Q_{\downarrow}$ and can be parameterized as:
\begin{equation} \label{eq:Jdia}
    J'_\mathrm{dia} \propto \delta Q_{\downarrow}.
\end{equation}
Similarly, we define the equilibrium diabatic heating rate when the gradient of convective activity is absent, $J^{\#}_\mathrm{dia}$. The quantity $J^{\#}_\mathrm{dia}$ is also proportional to $\mathrm{PW'}$, reflecting the well-known relation that diabatic heating is stronger when the air column is moister \citep{bretherton2004new}. How $J^{\#}_\mathrm{dia}$ depends on $\mathrm{PW'}$ also lacks an analytical theory. Nevertheless, given that robust semi-empirical relations have already been proposed \cite[e.g.,][]{fuchs2002GW,bretherton2004new}, we leave this topic for future work and focus on the nonlocal factor $l$ that influences the vortex size. Substituting Eq. (\ref{eq:Jdia}) into Eq. (\ref{eq:reaction_diffusion_1D}), we obtain:  
\begin{equation}  \label{eq:J_reaction_diffusion}
    -l^2 \frac{\partial^2}{\partial x^2} J'_\mathrm{dia} + J'_\mathrm{dia} = J^{\#}_\mathrm{dia}.
\end{equation}

\subsection{From 1-D to 2-D}

The cloud-chain model is derived in 1-D, yet the cloud-permitting simulations have a 2-D distribution of clouds. Appendix B extends the cloud chain model to 2-D, assuming that convective initiation is due to the collision of four gust fronts generated by the neighboring clouds:
\begin{equation}  \label{eq:J_2D_reaction_diffusion}
    -l^2 \left( \frac{\partial^2}{\partial x^2} + \frac{\partial^2}{\partial y^2} \right) J'_\mathrm{dia} + J'_\mathrm{dia} = J^{\#}_\mathrm{dia}.
\end{equation}
The infinite-domain Green's function of the 2-D problem is the zeroth-order Bessel function of the second kind, $K_0$:
\begin{equation} \label{eq:Bessel}
\begin{split}
     J'_\mathrm{dia} = \frac{1}{2\pi l^2} \iint\limits_{-\infty}^{\infty} K_0 \left(\frac{|\mathbf{x}-\mathbf{x''}|}{l} \right) J^{\#}_\mathrm{dia}(t,\mathbf{x''}) d\mathbf{x''}.
\end{split}    
\end{equation}

\subsection{Coupling the cloud chain model with the QG model}

\citet{fu2024small_part2} built a linear four-layer QG model and found that the problem can be approximately reduced to an effective single-layer case. Here, a heuristic derivation method is adopted that starts directly from a single-layer QG equation. The QG streamfunction of a single-layer system ($\psi$) obeys the potential vorticity equation \cite[e.g.,][]{vallis2017atmospheric}:
\begin{equation}  \label{eq:QG}
   \frac{\partial}{\partial t} \left[ \left( \frac{\partial^2}{\partial x^2} + \frac{\partial^2}{\partial y^2} \right) \psi - \frac{\psi}{L_R^2} \right] = \frac{f}{N^2}\frac{J'_\mathrm{dia}}{H},
\end{equation}
where $L_R = c/f$ is the Rossby deformation radius, $c$ is the effective gravity wave speed, $N$ is the buoyancy frequency, and $H$ is the effective thickness of the layer. The right-hand side is the potential vorticity production rate. The cloud chain model indicates that the anomalous diabatic heating rate $J'_\mathrm{dia}$ is a nonlocal function of $\mathrm{PW}'$, expressed as a convolution with the Bessel kernel:
\begin{equation}
\begin{split}
   J^{\#}_\mathrm{dia} 
   &=  \frac{1}{2\pi l^2} \iint\limits_{-\infty}^{\infty} K_0 \left(\frac{|\mathbf{x}-\mathbf{x''}|}{l} \right) \beta \mathrm{PW}' d\mathbf{x''},
\end{split}   
\end{equation}
where $\beta$ is a proportional factor. The $\mathrm{PW}'$ obeys:
\begin{equation}
    \frac{\partial \mathrm{PW'}}{\partial t} = - \frac{\mathrm{PW'}}{\tau} - \gamma \delta,
\end{equation}
where $\delta$ is the divergence in the layer, $\gamma$ represents the lapse rate of water vapor mixing ratio, and $\tau$ is the water vapor removal timescale due to precipitation. The divergence can be inferred from $\psi$ using the linearized vertical vorticity equation:
\begin{equation}  \label{eq:vort}
    \frac{\partial}{\partial t} \underbrace{ \left( \frac{\partial^2}{\partial x^2} + \frac{\partial^2}{\partial y^2} \right) \psi}_{\mathrm{vorticity}} = - f \delta.
\end{equation}
Equations (\ref{eq:QG})-(\ref{eq:vort}) form a closed equation set for studying the growth of a small-amplitude mesoscale perturbation. The instability mechanism is sketched here: a positive humidity perturbation induces greater diabatic heating, producing a relative vorticity anomaly. Such an anomaly arises from the stretching of planetary vorticity and, therefore, a convergent flow.  

The exponential growth rate of the system is defined as $\sigma$. Substituting in a normal mode solution in the form of $\exp \left[\sigma t + i (k_x x + k_y y) \right]$, we obtain the spectral growth rate:
\begin{equation}  \label{eq:sigma}
    \sigma = \frac{\beta \gamma}{N^2 H}
    \underbrace{ \left( 1 + \frac{1}{K^2 L_R^2} \right)^{-1} }_{\text{planet rotation}}
    \underbrace{\left( 1+K^2 l^2 \right)^{-1}}_{\text{cold pool}} - \frac{1}{\tau}.
\end{equation}
where $K \equiv \sqrt{k_x^2+k_y^2}$. Equation (\ref{eq:sigma}) takes the same form as the result of the four-layer QG model of \citet{fu2024small_part2}. The effective gravity wave speed calculated from the four-layer QG model is $c \approx 8$ m s$^{-1}$. Such a small $c$ value is mainly attributed to the zigzagged vertical profile of diabatic heating, which results from the coexistence of shallow, deep, and stratiform clouds. Other factors include the relatively weak stratification at the lower and upper levels, as well as the influence of condensation on the effective stratification.

The $K$ where $\sigma$ takes the maximum value is referred to as the most unstable wavelength. This wave mode grows faster than other modes, setting the size of TC embryo vortices. When there is a scale separation between the cold pool influence and the planetary rotation influence ($l \ll L_R$), the growth rate near its peak can be approximated as:
\begin{equation}
    \sigma \approx \frac{\beta \gamma}{N^2 H} \left( 1 - \frac{1}{K^2 L_R^2} \right) \left( 1 - K^2 l^2 \right) - \frac{1}{\tau}.
\end{equation}
Calculating $\partial \sigma/\partial K = 0$ yields the most unstable wavelength $\lambda_m$, which has a simple expression:
\begin{equation}  \label{eq:lamda_m_theory}
\begin{split}
    \lambda_m 
    &\approx 2\pi \left( l L_R \right)^{1/2} \\
    &= 2\pi \left( \frac{l c}{f} \right)^{1/2}.
\end{split}    
\end{equation}
This indicates that the vortex size depends on the geometric average of $l$ and $L_R$. A higher $l$ or smaller $f$ increases the vortex size, consistent with the CM1 simulations (Fig. \ref{fig:spectrum}). 

The four length scales involved in the TC embryo formation are summarized in Table \ref{Table_length}: the cold pool radius $R$, the nonlocal influence length scale due to cold pools $l$, the TC embryo size $\lambda_m$, and the Rossby deformation radius $L_R$. They obey:
\begin{equation}
    R \lesssim l \ll \lambda_m \lesssim L_R.
\end{equation}

\begin{table}[h]
\caption{The four basic length scales involved in the TC embryo size theory. }\label{Table_length}
\begin{center}
\begin{tabular}{ccccrrcrc}
\hline
\hline
Name  &  Meaning &  Diagnostic method & Modeling method \\
\hline
$R$   & Mean cold pool radius & Autocorrelation of $q'_s$ (Appendix A) & Survival competition model \citep{fu2025mean_part1,fu2025mean_part2}\\
$l$   & Influence length of PW & Not directly diagnosed & Cloud chain model \\
$\lambda_m$  & TC embryo size & Spectrum of PW (section \ref{sec:CM1}) & Linear stability analysis \\
$L_R$  & Rossby deformation radius & Not directly diagnosed & Four-layer QG model \citep{fu2024small_part2} \\
\hline
\hline
\end{tabular}
\end{center}
\end{table}

\subsection{Validating the TC embryo size theory}

\begin{figure}[h] \centerline{\includegraphics[width=0.95\linewidth]{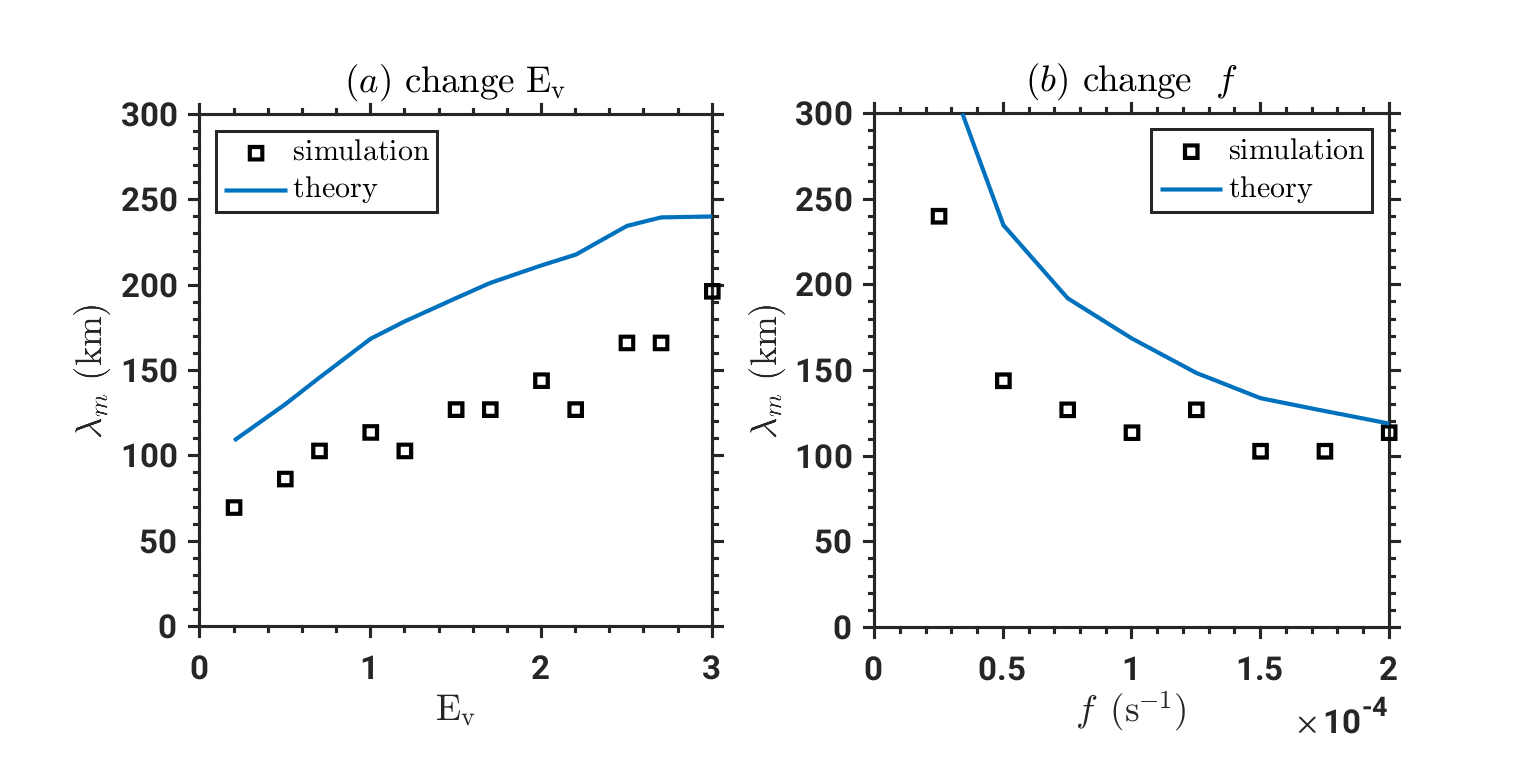}}
\caption{ Benchmark of the TC embryo size theory with the cloud-permitting simulations. The squares show the most unstable wavelength $\lambda_m$ diagnosed from cloud-permitting simulations, referring to Fig. \ref{fig:spectrum}. The maximization calculation is preconditioned by smoothing the spectral curves with a 10-point Gaussian filter. The solid line shows the theoretical prediction of $\lambda_m$, using Eq. (\ref{eq:lamda_m_theory}). (a) shows experiments with varying $\mathrm{E_v}$, and (b) shows experiments with varying $f$.}   \label{fig:theory}
\end{figure}

Figure \ref{fig:theory} plots the theoretically calculated $\lambda_m$. The theory captures the increase of vortex size with cold pool intensity ($\mathrm{E_v}$) and the decrease of vortex size with planetary rotation ($f$). Nevertheless, the $\lambda_m$ is systematically overestimated by 30\%. The author speculates that a significant uncertainty lies in $\alpha$, which represents the fractional contribution of the cold pool to convective initiation. The current diagnosis of $\alpha$ is too crude. It is based on the mean kinetic energy of the near-surface irrotational wind and the standard deviation of near-surface potential temperature. The influence of CIN on convective initiation may be parameterized as a reduction factor applied to $\alpha$, thereby explaining the gap between theory and simulations. Despite the discrepancy, the theory has some predictive power for TC embryo size in cloud-permitting simulations.

\section{Discussion}\label{sec:discussion}

This section reflects on the limitations of the cloud chain model and outlines potential future directions. 

First, the cloud-chain model only works for a weakly non-equilibrium state. The current model formulation, even for the fully nonlinear model [Eq. (\ref{eq:nondim_chain})], requires the cold pools to be negatively buoyant when colliding. This presents a problem when studying the dynamics at the edge of a mesoscale cold pool \citep{windmiller2019convection,yanase2020new}, where no cold pool is coming in from outside. In this case, the equation would still calculate the surface sensible heating and report a positive $\theta'^n_{\uparrow,j\pm 1}$. As a result, the mechanical lifting part, $(-\theta'^n_{\uparrow,j\pm 1})^{1/2}$, would give an imaginary number. One possible way to resolve this issue is to reset $\theta'^n_{\uparrow,j\pm 1}$ to zero if it is positive. This revision may extend the model to predict the evolution of a more intense vortex, such as a tropical depression.

Second, the geometry of the cloud chain model is much simpler than that of the cloud population in the cloud-permitting simulations and in the real atmosphere. The model has been extended to a 2-D Cartesian cloud chain in Appendix B, but the pattern is still too simple. For future work, extensions to a hexagonal mesh \cite[e.g.,][]{asai1967cloud}, or a triangular mesh \cite[e.g.,][]{meyer2020mechanical} can be considered. In addition, cold pool collision involves many stochastic factors that are hard to handle with a fixed mesh \citep[e.g.,][]{haerter2019circling,nissen2021circling}. Future modeling work may add noise to the cloud position and test its influence on TC embryo size prediction. 

%These extensions would help answer whether the geometric pattern of clouds matters for the theoretical prediction of $l$. 

%The convection initiated by cold pools with unequal strength and even a single cold pool also need to be considered \cite[e.g.,][]{feng2015cold_pool,van2017collision,yang2021parcel}. 

Third, the influence of vertical shear on convective initiation \citep{rotunno1988theory,davis2015formation} needs to be considered. The down-shear enhancement effect may add an ``effective drift velocity" to the damping-diffusion equation of cloud interaction [Eq. (\ref{eq:reaction_diffusion_1D})]. Whether the drift velocity favors or disfavors convective organization can be investigated. In addition, a stronger vertical shear may lengthen the convective phase of the lifecycle \cite[e.g.,][]{emanuel1994book}. This would break the assumption that the convective phase is much shorter than the cold pool phase, which is marginally satisfied ($1$ hour vs. $2.5$ hours scale) even if shear is absent. These extensions are crucial for developing a model for TC embryo evolution in a vertically sheared flow \cite[e.g.,][]{nam2023bifurcation}.   

Fourth, some incompleteness of the theory is highlighted. The water vapor mixing ratio at the gust front (cold pool edge) is modeled using the bulk model, which is more suitable for an averaged quantity in a cold pool than a local quantity. The partition between cold pool and convective cells in initiating convection, as represented by $\alpha$, is diagnosed rather crudely. The influence of CIN on convective initiation also needs to be considered.  

%is the only tuning parameter of the theory. Using $XXX=2$ or 2.5 yields a good agreement with the cloud-permitting simulation, but it comes from parameter tuning and therefore should be viewed cautiously. The other key input quantity to the theory is the rain-surface evaporation ratio $\mathcal{R}_q$. Theoretically speaking, $\mathcal{R}_q$ depends on the terminal cold pool radius $R$, but for simplicity, we choose to diagnose it. The diagnostic method of $\mathcal{R}_q$, which picks out cold pool regions by subtracting a domain minimum value, is also crude and can be updated in the future.  

\section{Conclusion}  \label{sec:conclusion_wavepacket}

Cold pools are known as a medium that spreads convective activity and influences the size of a TC embryo vortex. Still, theoretical modeling of the cold pool effect remains challenging. The key underlying question is: how does the intensity of convection depend on the cold pools initiating it and, ultimately, the intensity of neighboring convection? 

This paper investigates this problem through cloud-permitting simulations of spontaneously generated TC embryo vortices, focusing on the first few days when the convective activity yields a weak deviation from the homogeneous (equilibrium) state. In the simulations, the sub-cloud evaporation rate in the \citet{morrison2005new} microphysics scheme has been multiplied by a factor of $\mathrm{E_v}$. Stronger sub-cloud evaporation produces wider mesoscale vortices that grow more slowly, indicating a ``diffusive" process.

A chain model with uniform cloud spacing is built to study the mechanism of diffusion, specifically how an anomalously humid air column influences distant convective activity. The chain evolves alternatively between a convective phase and a cold pool phase. During the convective phase, sub-cloud moisture convergence determines the updraft and downdraft intensities and, therefore, the cold pool's initial potential temperature and humidity. During the cold pool phase, the cold pool gets warmed and humidified by the surface sensible and latent heating. The cold pool's potential temperature deficit indicates its collisional speed. The cold pools and mixed-layer convective cells together determine the sub-cloud convergence wind speed, referred to as the mechanical lifting effect, as well as the water-vapor anomaly, referred to as the thermodynamic forcing effect.

A small perturbation analysis of the cloud chain model yields a damping-diffusion equation of length scale $l$, which shows how far a locally humid air column influences the distant convective activity before it is damped out. The influence length scale depends on the mean cold pool radius, $R$, and a convective memory weight, $\mathcal{W}$. The $\mathcal{W}$ is contributed by the sum of the mechanical lifting component ($\mathcal{W}_{\theta}$) and the thermodynamic forcing component ($\mathcal{W}_{q}$), multiplied by a factor of $\alpha$ that depicts the fractional contribution of cold pools to convective initiation. The $\alpha$ drops as rain evaporation weakens. The $\mathcal{W}_{\theta}$ and $\mathcal{W}_{q}$ have the following properties:
\begin{itemize}
    \item For mechanical lifting, stronger rain evaporation reduces $\mathcal{W}_{\theta}$. Surface sensible heating is an external contribution that does not carry the information of the previous convection. Stronger rain evaporative cooling reduces the importance of surface sensible heating, thereby increasing the buoyant deficit of the gust front and, consequently, making convective initiation harder. Thus, stronger rain evaporative cooling ``devalues" itself, making mechanical lifting less sensitive to it and reducing memory.
    \item For thermodynamic forcing, stronger rain evaporation increases $\mathcal{W}_q$. Both rain evaporation and surface evaporation moisten the gust front, but only rain evaporation carries the information of the previous convection. Thus, stronger rain evaporation makes the humidity memory from neighboring clouds more significant. 
\end{itemize}

%For the more general case, a correction factor $\alpha$ needs to be multiplied on $\mathcal{W}_{\theta}$ and . The $\alpha$ depicts the fractional contribution of cold pools to convective initiation. It drops as rain evaporation weakens. 

A model of TC embryo evolution is built by coupling the cloud chain model with the quasi-geostrophic (QG) model. The most unstable wavelength $\lambda_m$, which represents the spontaneously generated TC embryo size, is predicted to be set by the geometric average of the nonlocal influence length scale $l$ and the Rossby deformation radius. The $\lambda_m$ increases with the rain evaporation rate and decreases with the planetary rotation rate. The trend is in good agreement with cloud-permitting simulations, but the magnitude of the TC embryo size is overestimated by 30\%.

Despite several limitations, this paper derives an analytical model of TC embryo vortex size, built on a theoretical framework for the non-equilibrium aspects of cloud interaction. This framework dives into the details of cold pool evolution, providing a physical basis for the hypothesis raised by previous studies: cloud interaction may be depicted with a damping-diffusion type equation or a nonlocal smoother, which laterally expands the convective system \cite[e.g.,][]{craig2013coarsening,hottovy2015spatiotemporal,ahmed2019stochastic,windmiller2019universality,biagioli2023dimensionless,fu2025quasi,yanase2025nonlocal}.

%%%%%%%%%%%%%%%%%%%%%%%%%%%%%%%%%%%%%%%%%%%%%%%%%%%%%%%%%%%%%%%%%%%%%
% ACKNOWLEDGMENTS
%%%%%%%%%%%%%%%%%%%%%%%%%%%%%%%%%%%%%%%%%%%%%%%%%%%%%%%%%%%%%%%%%%%%%
\acknowledgments
This curiosity-driven research spanned four years, during which the author received almost equal support from three institutes. Some preliminary theoretical results were reported in Chapter 5 of the author's Stanford University Ph.D. thesis. The initial theory validation and the first draft were made during the author's stay at the University of Chicago, supported by the T. C. Chamberlin Postdoctoral Fellowship. A substantial improvement to the model framework and interpretation was made after the author arrived at Nanjing University, supported by faculty startup funding No. 1480604113. We thank Dr. Morgan O'Neill at The University of Toronto, Dr. Da Yang at Stanford University, Dr. PJ Tuckman at the University of Chicago, Dr. Zhaohua Wu at Florida State University, Dr. David Nolan at the University of Miami, Dr. Junfei Li at Purdue University, and Mr. Yue Shen at Nanjing University for inspiring discussions. We thank George Bryan at NCAR for developing and maintaining the Bryan Cloud Model 1. We thank Stanford Research Computing Center and NCAR CISL Lab (Derecho cluster) for providing computational resources. Last but not least, we thank three anonymous reviewers for their insightful comments, which make the model framework cleaner and more rigorous.  

%%%%%%%%%%%%%%%%%%%%%%%%%%%%%%%%%%%%%%%%%%%%%%%%%%%%%%%%%%%%%%%%%%%%%
% DATA AVAILABILITY STATEMENT
%%%%%%%%%%%%%%%%%%%%%%%%%%%%%%%%%%%%%%%%%%%%%%%%%%%%%%%%%%%%%%%%%%%%%
% 
%
\datastatement
The movie version of Fig. \ref{fig:pcolor}, two supplementary figures, a math derivation note, and the postprocessing codes are deposited on Zenodo: https://doi.org/10.5281/zenodo.17668821. The CM1 simulation data can be obtained by contacting the author.

%%%%%%%%%%%%%%%%%%%%%%%%%%%%%%%%%%%%%%%%%%%%%%%%%%%%%%%%%%%%%%%%%%%%%
% APPENDIXES
%%%%%%%%%%%%%%%%%%%%%%%%%%%%%%%%%%%%%%%%%%%%%%%%%%%%%%%%%%%%%%%%%%%%%

\appendix[A]
\appendixtitle{Diagnosing the mean cold pool radius}

\begin{figure}[h]
\centerline{\includegraphics[width=1\linewidth]{FigA1_R_array.png}}
  \caption{(a) The 1-D spatial autocorrelation function of near-surface ($z=25$ m) water vapor mixing ratio in the cloud-permitting simulations, using the data at $t=2.5$ days. The solid lines from dark to bright denote experiments with increasing $\mathrm{E_v}$. The dashed black line denotes the 0.5 threshold level for calculating $R$. (b) The diagnosed mean cold pool radius $R$ versus $\mathrm{E_v}$. The circle line denotes the threshold-based method, and the crosses denote the minimum-based method. }\label{fig:R_array}
\end{figure}   

The Fourier spectral diagnosis of cold pool size has a limitation in the weak-cold-pool regime ($\mathrm{E_v}\ll 1$). That is, the rapid growth of mesoscale vortices causes cold pools to vanish in relatively dry regions. Therefore, there is no peak wavelength for the near-surface ($z=25$ m) water vapor mixing ratio ($q_s$). To avoid this issue, this appendix uses the autocorrelation function of $q_s$, where the cold pool size manifests as the decorrelation length scale. The autocorrelation function is calculated as the average of two 1-D autocorrelation functions obtained by shifting the $q_s$ array in the x- and y-directions, respectively. 

The mean cold pool size can be straightforwardly identified as the length scale at which the autocorrelation function attains its minimum \citep{haerter2017onset}. Nevertheless, this lowest correlation length has a similar physical meaning to the peak wavelength in the spectral analysis, and the lowest autocorrelation length also does not exist for $\mathrm{E_v}\ll 1$ experiments. An alternative method is to diagnose the lag distance at which the autocorrelation drops to half (0.5), and then relate it to the mean cold pool size by multiplying a proportional coefficient. The two methods should yield the same result in the $\mathrm{E_v}\gg 1$ regime where the autocorrelation function has a clear minimum. This alignment property provides a calibration method, rendering a proportional coefficient of 3.2 that is to be multiplied by the ``0.5-threshold" lag distance (Fig. \ref{fig:R_array}b).

%This diagnostic method is the same as that used by \citet{fu2025mean_part1}, except that they employed the water vapor mixing ratio averaged vertically in the mixed layer, which may be overly complicated. 

%We have tried to use the data at $t=4$ days + 1 hour, the time used by all other diagnoses in the paper. However, the mixed layer is already significantly influenced by the inhomogeneity of convective activity, making the diagnostic result of $l$ in the $\mathrm{E_v}=0.2$ case (the fastest growing case) much higher than others. Thus, we use the $t=2.5$ days data instead, by which the convective activity is still relatively homogeneous. 

Figure \ref{fig:R_array} shows that $R$ monotonically increases with $\mathrm{E_v}$ from around 7.5 km to 10.5 km across the twelve experiments with varying $\mathrm{E_v}$. This diagnosis is consistent with the recently published cold pool size theory, derived from a combination of energy balance and survival competition among the cold pool population \citep{fu2025mean_part1,fu2025mean_part2}. Readers may wonder why $R$ seems not to drop to zero as $\mathrm{E_v} \to 0$. The author speculates that the error in simulating the cold pool size becomes significant as it approaches the horizontal grid spacing of 2 km. Experiments with a finer resolution is needed to confirm the trend of $R$ in the $\mathrm{E_v} \ll 1$ regime.

\appendix[B]
\appendixtitle{Extension to a 2-D cloud chain model}

This appendix extends the 1-D cloud chain model to 2-D. The 2-D model assumes clouds are distributed on a Cartesian mesh with indices $i$ and $j$. The convective initiation involves four gust fronts emitted by the four neighboring clouds: at $(i-1,j)$, $(i+1,j)$, $(i,j-1)$, $(i,j+1)$. The 1-D fully nonlinear cloud chain equation (\ref{eq:nondim_chain}) is extended to:
\begin{equation}  \label{eq:nondim_chain_2D}
\begin{split}
     \frac{\delta Q^{n+1}_{\downarrow,i,j}}{\langle Q_{\downarrow} \rangle} &= \left( 1 + \frac{\mathrm{PW}'_{i,j}}{\mathrm{PW}_*} \right) \times \frac{1}{4} 
    \sum_{k=-1,1} \sum_{p=-1,1}  \\ &\left\{ \left\{ 1 + \alpha \left[ \left( 1 + \frac{1}{1-1/\mathcal{R}_\theta} \frac{\delta Q^n_{\downarrow,i+k,j+p}}{\langle Q_\downarrow \rangle} \right)^{1/2} - 1 \right] \right\} \left\{ 1 + \frac{\alpha}{1+1/\mathcal{R}_q} \frac{\delta Q^n_{\downarrow,i+k,j+p}}{\langle Q_\downarrow \rangle} \right\} - 1 \right\}.
\end{split}    
\end{equation}
Here, $\sum \sum$ denotes the summation over the four neighboring points. The 1-D small perturbation equation (\ref{eq:linear_chain_down}) is extended to:
\begin{equation}  \label{eq:linear_chain_down_2D}
    \frac{\delta Q^n_{\downarrow,i,j}}{ \langle Q_{\downarrow} \rangle} \approx 
    \frac{\mathcal{W}}{4} 
    \left( \frac{\delta Q^n_{\downarrow,i-1,j}}{\langle Q_{\downarrow}\rangle} 
    + 
    \frac{\delta Q^n_{\downarrow,i+1,j}}{\langle Q_{\downarrow}\rangle}
    + \frac{\delta Q^n_{\downarrow,i,j-1}}{\langle Q_{\downarrow}\rangle}
    + \frac{\delta Q^n_{\downarrow,i,j+1}}{\langle Q_{\downarrow}\rangle}    
    \right) 
    + \frac{\mathrm{PW'}_{i,j}}{\mathrm{PW}_*},
\end{equation}
where $\mathcal{W}$ has the same expression as Eq. (\ref{eq:chain_weight}). Viewing the small perturbation equation as a finite-difference approximation to a differential equation, neglecting the time tendency term, we obtain the 2-D version of the damping-diffusion equation [Eq. (\ref{eq:reaction_diffusion_1D})]:
\begin{equation}
\label{eq:reaction_diffusion_2D}
    - l^2 \left( \frac{\partial^2}{\partial x^2} + \frac{\partial^2}{\partial y^2} \right) \delta Q_{\downarrow} + \delta Q_{\downarrow} = \delta Q^{\#}_{\downarrow},
\end{equation}
Equations (\ref{eq:reaction_diffusion_1D}) and (\ref{eq:reaction_diffusion_2D}) show that the governing equations of the 1-D and 2-D cloud chain are almost identical. However, their Green's functions differ. The additional geometric spreading effect in the 2-D problem leads to a more steeply decaying influence than the 1-D problem.

\appendix[C]
\appendixtitle{Derivation of the 1-D cloud chain model}

The sub-cloud moisture convergence equation (\ref{eq:moisture_convergence}), the gust front momentum balance (\ref{eq:ustar}), the gust front potential temperature equation (\ref{eq:theta_solution}), the gust front water vapor mixing ratio equation (\ref{eq:q_solution}), and enthalpy conservation relation in rain evaporation (\ref{eq:enthalpy}), and the conversion ratio of sub-cloud evaporation to sub-cloud moisture convergence (\ref{eq:eta}), yield a closed equation set for the cloud chain evolution. No approximations are used in this appendix. 

All variables are expressed as the sum of the equilibrium and perturbation components [Eq. (\ref{eq:equilibrium_decompose})]. We start by decomposing the moisture convergence equation (\ref{eq:moisture_convergence}):
\begin{equation}  \label{eq:Q_full}
    \langle Q_{\uparrow} \rangle + \delta Q^{n+1}_{\uparrow,j}
    = \rho L_v \tau_\uparrow S \times \frac{1}{2}\sum_{k=-1,1} \left[ \left( u_* + \langle u_\uparrow \rangle + \delta u^n_{\uparrow,j+k} \right) \left( q_* + \langle q'_\uparrow \rangle + \delta q'^n_{\uparrow,j+k} \right)\right].
\end{equation}
The equilibrium component is:
\begin{equation}  \label{eq:Q_equilibrium}
    \langle Q_{\uparrow} \rangle
    = \rho L_v \tau_\uparrow S   \left( u_* + \langle u_\uparrow \rangle \right) \left( q_* + \langle q'_\uparrow \rangle \right).
\end{equation}
Subtracting Eq. (\ref{eq:Q_equilibrium}) from Eq. (\ref{eq:Q_full}), we obtain the perturbation equation of sub-cloud moisture convergence:
\begin{equation}  \label{eq:Q_perturbation}
\begin{split}
    \delta Q^{n+1}_{\uparrow,j} 
    &= \rho L_v \tau_\uparrow S \left( u_* + \langle u_\uparrow \rangle \right) \left( q_* + \langle q'_\uparrow \rangle \right) \\
    &\times \frac{1}{2}\sum_{k=-1,1} \left\{ \left\{ 1 + \frac{\langle u_{\uparrow} \rangle}{u_* + \langle u_{\uparrow} \rangle} \frac{\delta u^n_{\uparrow,j+k}}{\langle u_\uparrow \rangle} \right\} \left\{ 1 +  \frac{\langle q_{\uparrow} \rangle}{q_* + \langle q_{\uparrow} \rangle} \frac{\delta q'^n_{\uparrow,j+k}}{\langle q'_\uparrow \rangle} \right\} - 1 \right\} \\
    &=\langle Q_{\uparrow} \rangle \times \frac{1}{2}\sum_{k=-1,1} \left\{ \left\{ 1 + \frac{\langle u_{\uparrow} \rangle}{u_* + \langle u_{\uparrow} \rangle} \frac{\delta u^n_{\uparrow,j+k}}{\langle u_\uparrow \rangle} \right\} \left\{ 1 +  \frac{\langle q_{\uparrow} \rangle}{q_* + \langle q_{\uparrow} \rangle} \frac{\delta q'^n_{\uparrow,j+k}}{\langle q'_\uparrow \rangle} \right\} - 1 \right\}.
\end{split}
\end{equation}
The convergence wind speed can be expressed in terms of potential temperature using the gust front momentum balance [Eq. (\ref{eq:ustar})]. The decomposed version of Eq. (\ref{eq:ustar}) is:
\begin{equation}
    \langle u_{\uparrow} \rangle + \delta u^n_{\uparrow,j\pm1} = \mathrm{Fr} \left( - g H_f \frac{\langle \theta_{\uparrow} \rangle + \delta \theta'^n_{\uparrow,j\pm1}}{\overline{\theta}} \right)^{1/2}.
\end{equation}
Its equilibrium and perturbation components are:
\begin{equation}  \label{eq:u_perturbation}
\begin{split}
\langle u_{\uparrow} \rangle &= \mathrm{Fr} \left( - g H_f \frac{\langle \theta_{\uparrow} \rangle}{\overline{\theta}} \right)^{1/2},\\
\delta u^n_{\uparrow,j\pm1} &= \langle u_{\uparrow} \rangle \left[ \left( 1 + \frac{\delta \theta'^n_{\uparrow,j\pm1}}{\langle \theta_{\uparrow} \rangle} \right)^{1/2} - 1 \right].    
\end{split}    
\end{equation}
Substituting Eq. (\ref{eq:u_perturbation}) into (\ref{eq:Q_perturbation}), we obtain:
\begin{equation}  \label{eq:Q_simplified}
\begin{split}
    \frac{\delta Q^{n+1}_{\uparrow,j}}{\langle Q_{\uparrow} \rangle} =  \frac{1}{2}\sum_{k=-1,1} \left\{ \left\{ 1 + \frac{\langle u_{\uparrow} \rangle}{u_* + \langle u_{\uparrow} \rangle} \left[ \left( 1 + \frac{\delta \theta'^n_{\uparrow,j+k}}{\langle \theta'_\uparrow \rangle} \right)^{1/2} - 1 \right] \right\} \left\{ 1 +  \frac{\langle q_{\uparrow} \rangle}{q_* + \langle q_{\uparrow} \rangle} \frac{\delta q'^n_{\uparrow,j+k}}{\langle q'_\uparrow \rangle} \right\} - 1 \right\}.    
\end{split}    
\end{equation}
Using Eq. (\ref{eq:alpha}) to express the fractional contribution of cold pools to sub-cloud moisture convergence ($\alpha$) in Eq. (\ref{eq:Q_simplified}), we obtain:
\begin{equation}
\label{eq:Q_simplified_alpha}
    \frac{\delta Q^{n+1}_{\uparrow,j}}{\langle Q_{\uparrow} \rangle} =
    \frac{1}{2}\sum_{k=-1,1} \left\{ \left\{ 1 + \alpha \left[ \left( 1 + \frac{\delta \theta'^n_{\uparrow,j+k}}{\langle \theta'_\uparrow \rangle} \right)^{1/2} - 1 \right] \right\} \left\{ 1 + \alpha \frac{\delta q'^n_{\uparrow,j+k}}{\langle q'_\uparrow \rangle} \right\} - 1 \right\}.   
\end{equation}

Next, we express $\delta Q^{n+1}_{\uparrow,j}$, $\delta \theta'^{n}_{\uparrow,j}$, and $\delta q'^{n}_{\uparrow,j}$ with the sub-cloud rain evaporation amount per cold pool $\delta Q^{n}_{\downarrow,j\pm1}$. The $\delta Q^{n+1}_{\uparrow,j}$ can be first expressed with $\delta Q^{n+1}_{\downarrow,j}$ using the decomposed version of the conversion ratio equation (\ref{eq:eta}):
\begin{equation}
    \langle Q_{\downarrow} \rangle + \delta Q^{n+1}_{\downarrow,j} = \eta_* \left( 1 + \frac{\mathrm{PW}'_j}{\mathrm{PW}_*} \right) \left( \langle Q_{\uparrow} \rangle + \delta Q^{n+1}_{\uparrow,j} \right),
\end{equation}
which represents how the air-column humidity modifies the sub-cloud rain evaporation. Its equilibrium and perturbation components are:
\begin{equation} \label{eq:eta_perturbation}
\begin{split}
    \langle Q_{\downarrow} \rangle &= \eta_* \langle Q_{\uparrow} \rangle, \\
    \delta Q^{n+1}_{\downarrow,j} &= \eta_* \delta Q^n_{\uparrow,j} + \frac{\mathrm{PW}'^n_j}{\mathrm{PW}_*} \langle Q_{\uparrow} \rangle.
\end{split}
\end{equation}
The decomposed version of the gust front potential temperature equation (\ref{eq:theta_solution}) is:
\begin{equation}   \label{eq:theta_full}
\begin{split}
    \langle \theta'_{\uparrow} \rangle + \delta \theta'^n_{\uparrow,j}  
    &= \langle \theta'_{\downarrow} \rangle \, e^{-\varepsilon {R}}
    +  \delta \theta'^n_{\downarrow,j} \, e^{-\varepsilon {R}}  
    + 
    \frac{2\pi}{9}\frac{{C} \Delta \theta}{{V}} {R}^3\,e^{-\varepsilon {R}}.
\end{split}    
\end{equation}
Its equilibrium and perturbation components are:
\begin{equation}   \label{eq:theta_perturbation}
\begin{split}
    \langle \theta'_{\uparrow} \rangle
    &= \langle \theta'_{\downarrow} \rangle \, e^{-\varepsilon {R}}
    + 
    \frac{2\pi}{9}\frac{{C} \Delta \theta}{{V}} {R}^3\,e^{-\varepsilon {R}}, \\
    \delta \theta'^n_{\uparrow,j}  &= \delta \theta'^n_{\downarrow,j} \, e^{-\varepsilon {R}}. 
\end{split}    
\end{equation}
The decomposed version of the gust front water vapor mixing ratio equation (\ref{eq:q_solution}) is:
\begin{equation}   \label{eq:q_full}
\begin{split}
    \langle q'_{\uparrow} \rangle + \delta q'^n_{\uparrow,j}  
    &= \langle q'_{\downarrow} \rangle \, e^{-\varepsilon {R}}
    +  \delta q'^n_{\downarrow,j} \, e^{-\varepsilon {R}}  
    + 
    \frac{2\pi}{9}\frac{{C} \Delta q}{{V}} {R}^3\,e^{-\varepsilon {R}}.
\end{split}    
\end{equation}
Its equilibrium and perturbation components are:
\begin{equation}   \label{eq:q_perturbation}
\begin{split}
    \langle q'_{\uparrow} \rangle
    &= \langle q'_{\downarrow} \rangle \, e^{-\varepsilon {R}}
    + 
    \frac{2\pi}{9}\frac{{C} \Delta q}{{V}} {R}^3\,e^{-\varepsilon {R}}, \\
    \delta q'^n_{\uparrow,j}  &= \delta q'^n_{\downarrow,j} \, e^{-\varepsilon {R}}. 
\end{split}    
\end{equation}
Equations (\ref{eq:theta_perturbation}) and (\ref{eq:q_perturbation}) yield the relation between the gust front properties upon formation and upon collision:
\begin{equation}  \label{eq:up_down_theta}
    \frac{\delta \theta'^n_{\uparrow,j}}{\langle \theta'_{\uparrow} \rangle} = \frac{\delta \theta'^n_{\downarrow,j}}{\langle \theta'_{\downarrow} \rangle} \frac{1}{1-1/\mathcal{R}_\theta},
\end{equation}
\begin{equation}
\label{eq:up_down_q}
    \frac{\delta q'^n_{\uparrow,j}}{\langle q'_{\uparrow} \rangle} = \frac{\delta q'^n_{\downarrow,j}}{\langle q'_{\downarrow} \rangle} \frac{1}{1+1/\mathcal{R}_q}.
\end{equation}
The signs in front of $1/\mathcal{R}_\theta$ and $1/\mathcal{R}_q$ are different because the surface sensible heat flux and rain evaporative cooling have opposite effects, and the surface evaporative flux and rain evaporative moistening have aligned effects. The enthalpy conservation relation in rain evaporation [Eq. (\ref{eq:enthalpy})] relates $\delta Q^{n+1}_{\downarrow,j}$, $\delta \theta'^{n+1}_{\downarrow,j}$, and $\delta q'^{n+1}_{\downarrow,j}$ with:
\begin{equation}  \label{eq:enthalpy_perturbation}
    \frac{\delta Q^{n+1}_{\downarrow,j}}{\langle Q_{\downarrow} \rangle}= \frac{\delta \theta'^{n+1}_{\downarrow,j}}{\langle \theta'_{\downarrow} \rangle}=\frac{\delta q'^{n+1}_{\downarrow,j}}{\langle q'_{\downarrow} \rangle}.
\end{equation}
Substituting Eqs. (\ref{eq:eta_perturbation}) and (\ref{eq:up_down_theta})-(\ref{eq:enthalpy_perturbation}) into Eq. (\ref{eq:Q_simplified_alpha}), we obtain the cloud chain model with the perturbation component of the sub-cloud rain evaporation amount $\delta Q^n_{\downarrow,j}$ as the single dependent variable:
\begin{equation*}
\begin{split}
     \frac{\delta Q^{n+1}_{\downarrow,j}}{\langle Q_{\downarrow} \rangle} &= \left( 1 + \frac{\mathrm{PW}'_j}{\mathrm{PW}_*} \right) \\
     &\times \frac{1}{2}\sum_{k=-1,1} \left\{ \left\{ 1 + \alpha \left[ \left( 1 + \frac{1}{1-1/\mathcal{R}_\theta} \frac{\delta Q^n_{\downarrow,j+k}}{\langle Q_\downarrow \rangle} \right)^{1/2} - 1 \right] \right\} \left\{ 1 + \frac{\alpha}{1+1/\mathcal{R}_q} \frac{\delta Q^n_{\downarrow,j+k}}{\langle Q_\downarrow \rangle} \right\} - 1 \right\},
\end{split}
\end{equation*}
which serves as Eq. (\ref{eq:nondim_chain}). A small perturbation analysis in section \ref{sec:theory}\ref{subsec:small_perturbation} reduces the cloud chain model to a damping-diffusion equation. In section \ref{sec:validation}, it is coupled with the quasi-geostrophic equation to solve for the TC embryo size.

%%%%%%%%%%%%%%%%%%%%%%%%%%%%%%%%%%%%%%%%%%%%%%%%%%%%%%%%%%%%%%%%%%%%%
% REFERENCES
%%%%%%%%%%%%%%%%%%%%%%%%%%%%%%%%%%%%%%%%%%%%%%%%%%%%%%%%%%%%%%%%%%%%%

\bibliographystyle{ametsocV6}
\bibliography{references}

\end{document}